\documentclass[pre,twocolumn,showpacs]{revtex4}
\usepackage{graphicx}
\usepackage{amssymb}
\usepackage{longtable}
\begin{document}

\title{Characterizing large scale base composition structures of genomes}
\author{Zhengqing Ouyang$^{1}$, Jian Liu$^{1}$ and Zhen-Su
She$^{1,2,}\footnote{Electronic address: she@pku.edu.cn}$}
\affiliation{%
$^1$ State Key Lab for Turbulence and Complex Systems \\
and Center for Theoretical Biology, Peking University, Beijing 100871, P.R. China\\
$^2$ Department of Mathematics, UCLA, Los Angeles, CA 90095, USA}

\date{\today}

\begin{abstract}
Intermittent density fluctuations of nucleotide molecules
(adenine, guanine, cytosine and thymine) along DNA sequences are
studied in the framework of a hierarchical structure (HS) model
originally proposed for the study of fully developed turbulence
[She and L\'{e}v\^{e}que, Phys. Rev. Lett. {\bf 72}, 336 (1994)].
Large scale ($10^3 \le \ell \le 10^5$ bp) base density fluctuation
is shown to satisfy the HS similarity. The derived values of a HS
parameter $\beta$ from a large number of genome data (including
Bacteria, Archaea, human chromosomes and viruses) characterize
different biological properties such as strand symmetry,
phylogenetic relations and horizontal gene transfer. It is
suggested that the HS analysis offers a useful quantitative
description for heterogeneity, sequence complexity and large scale
structures of genomes.
\end{abstract}


\vspace{5pt}

\maketitle

\section{\label{sec:introduction}introduction}

The DNA sequence of a complete genome of an organism contains the
information not only for making all the proteins (genes) necessary
for the organism, but also for assembling them to form the
organism in a specific time order with specific three-dimensional
patterns. While small-scale (from several to hundreds base pairs)
patterns of the nucleotide arrangement are certainly important for
determining its coding or non-coding nature and some regulatory
biological functions (e.g. binding site or splicing site signal)
\cite{Boffelli04}, more large-scale variation across several
thousands base pairs or longer may be related to higher level
biological functions such as controlling networks of genes which
are likely important indices in evolution \cite{Banerjee02}. It is
important to develop tools for analyzing these patterns with the
available sequence data and to use it as a laboratory for
quantitative exploring biological laws such as the mechanism of
biological evolution \cite{Dujon04}.

There has been considerable efforts in studying the statistical
property of nucleotide distribution pattern \cite{Reddy95}. The
concept of ``domains-within-domains" has been introduced in
Ref.~\cite{Li94} and confirmed in Ref.~\cite{Bernaola96}. The
algorithms for DNA sequence alignment and similarity search have
been developed for the study of phylogeny and evolution of many
biological species \cite{Sugden03}. Other methods developed in
nonlinear analysis and information theory were introduced to
characterize coding and non-coding DNA sequences \cite{coding}.
Beside these studies focused on local motifs of the DNA sequence,
many other methods, including statistical physics analysis
\cite{Peng92}, spectrum analysis \cite{Li94,Voss92,Buldyrev95},
wavelet analysis \cite{Arneodo}, etc, have also been proposed to
measure the correlation between nucleotides over long distances
along one-dimensional DNA chain. Although long-range correlation
in DNA sequences has been established \cite{Li97}, the nature and
the significance of this correlative property remain under debate
\cite{long-corr}. Of previous analysis of interesting scaling
behaviors of DNA sequence, the most famous one is the $1/f$-like
power law at moderate length scales (typically $10-1000$ bp)
\cite{Voss92,Li92}. Less effort has been performed to examine
larger scale correlations, partially due to the lack of very long
sequences in the past decades. Recently, large scale structure of
DNA sequence, especially complete genomes, has been studied
\cite{Vieira99}. Such large scale structure at the genomic level
contains the global evolution information which is lack in small
scale one. Traditional methods, however, is ineffective to analyze
long range correlation structure at the whole genome level. For
example, local ``base-base" correlation is difficult to reveal the
corresponding biological meaning \cite{Li97}; power spectrum
analysis is impractical due to computer limitation
\cite{Vieira99}. The present study gives a different approach
which may be applicable to such large scale correlation at the
genome level.

We have briefly introduced a hierarchical structure (HS)
description of multiple scale structures of DNA sequences
\cite{Ouyang04} based on an earlier HS model for hydrodynamic
turbulence \cite{She94}. The starting point of the analysis is to
construct a nucleotide density fluctuation visualization and the
probability density function (PDF) description, then perform a
multiple moment scaling analysis \cite{ESS}, and apply the HS
scaling model to characterize the fluctuation structure. This
methodology makes it possible to study correlation structures up
to more than $10^5$ bp. Our analysis reveals that the nucleotide
composition variations along genomes are far from random, but
present a complex self-organized structure, called intermittent
structure, which can be captured by the HS analysis. In this work,
we will show that a detailed study of the systematic variation of
HS parameter $\beta$ measured from more than one hundred sequences
of four kingdoms (Bacteria, Archaea, human chromosomes and
viruses) reveals significantly biological information, such as
strand symmetry, phylogenetic relations and horizontal gene
transfer.

The paper is organized as follows: A multi-scale variable $f_\ell$
concerning base composition fluctuations of genome sequences is
introduced in Sec.~\ref{sec:base}. We present briefly measurements
of scaling property with special emphasis on the HS model and
similarity test ($\beta$-test) in Sec.~\ref{sec:hs}.
Section~\ref{sec:results} is devoted to a detailed HS analysis of
various kinds of genome data. Section~\ref{sec:con} offers a
summary and some additional discussion.

\section{\label{sec:base}Base composition fluctuations}

\begin{figure}
\includegraphics[width=9cm,height=10cm]{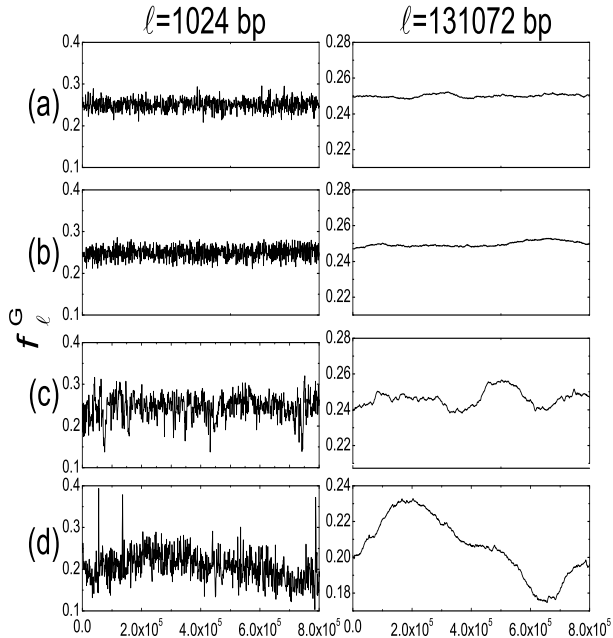}
\caption {\label{fig:fluc} Nucleotide guanine density (G)
variation $f_\ell^{G}$ of (a) Random, (b) Simulation, (c) Ecoli
and (d) Hsap4. Local densities are calculated over a window of
$\ell_{min}=2^{10}$ bp (the left) and $\ell_{max}=2^{17}$ bp (the
right), respectively. The sliding window moves at a step of length
$\Delta=1024$ bp. Note the intensive fluctuation of natural
sequence away from artificial ones.}
\end{figure}

\begin{figure}
\includegraphics[width=9cm]{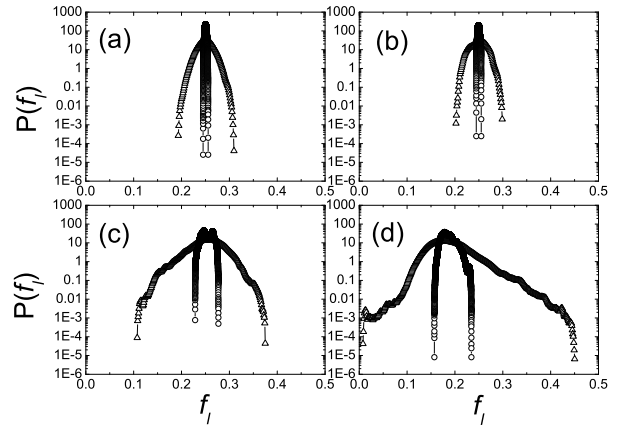}
\caption {\label{fig:pdf} Typical PDFs of guanine density (G)
$f_\ell$ of (a) Random, (b) Simulation, (c) Ecoli and (d) Hsap4 at
two scales $\ell_{min}=2^{10}$ (circle) and $\ell_{max}=2^{17}$
(triangle). Note that with $\ell$ decreasing, the right wings of
PDFs progress further, indicating the appearance of high intensity
fluctuations in $f_\ell$ which can be captured by the HS
analysis.}
\end{figure}

A single-stranded DNA chain can be viewed as a symbolic series $\{
n_i \} (i = 1, 2, ..., L)$ of length $L$ comprised of four
nucleotides A, C, G and T. There are many kinds of transformation
of DNA sequences to capturing certain properties, such as the
``DNA walk" \cite{Peng92}, which construct a numerical sequence
$\{ u_i \}$ by a certain mapping rule (e.g., adenine rule: if
$n_i=A$ then $u_i=1$; in all other cases $u_i=0$). Then a running
sum $y(n)=\sum_{i=1}^{n}[u_i-(1-u_i)]$ can be presented
graphically as a one-dimensional landscape of the original DNA
sequence. Here we employ an alternative approach that introduces a
window with length $\ell$ bp on the DNA sequence, and define the
(local) density of a particular base as
\begin{equation}
f_\ell=\frac{1}{\ell}\sum_{k=i}^{i+\ell-1}u_i,
\end{equation}
where $i$ is the location of the first base within the window. The
definition can be used to any single nucleotide (A, C, G, or T) or
their degeneracy (R, Y, etc.) and to any dinucleotide molecules
(AT, AG, etc.). By sliding the window with a certain moving step
$\Delta$ and changing the window size $\ell$ along the DNA
sequence, we can obtain different fluctuation sequences $f_\ell$.
When $\ell=L$, $f_\ell$ become the mean content of the nucleotide
in the entire DNA chain. This multi-scale variable $f_\ell$,
similar to the locally averaged energy dissipation rate
$\epsilon_\ell$ in the turbulence field, is the coarse gaining of
base density of the original DNA sequence, which allow us to
calculate the probability density functions (PDF) $P(f_\ell)$ and
other quantities of $f_\ell$ of interest.

The fluctuation structures of DNA sequences can be shown by a plot
of local base density $f_\ell$ against the sequence position.
Figure~\ref{fig:fluc} displays a segment of 0.8 million bp guanine
density (G) fluctuations $f_\ell$ with two scales $2^{10}$
($\approx 10^3$ bp) and $2^{17}$ ($\approx 10^5$ bp) for four
sequences: an independent identical distribution (\textit{i.i.d.})
random sequence with ten million bp and 50$\%$ A+T content
(Random), a simulated genome sequence with one million bp by the
minimal model (Simulation) \cite{Hsieh03}, \textit{E. coli} whole
genome (Ecoli) and \textit{H. sapiens} chromosome 4 contig 8
(Hsap4), respectively. The random sequence shows no surprising
white noise signal at both scales and has the least fluctuations
amplitude. The simulated sequence contains some visible tips but
in a whole is stationary. The \textit{E. coli} genome contains
many low guanine density region which is atypical to the main
body. The Hsap4 sequence with the special high guanine density
seems most intermittent, which include many strong bursts breaking
against the background and the highest fluctuation amplitude among
the four. We believe that the transition of the fluctuation from
small scales to large ones is of special interest to reveal the
global information of genome.

With the multi-scale variable $f_\ell$, we can carry out the
multi-scale PDF method to characterize the interesting structures
of such sequences. Firstly we perform an analytical discussion on
the random control sequence called \textit{i.i.d.}, i.e., the
probability $p$ of each base with $u_i=1$ (guanine rule) is equal
to 0.25. For a certain window size $\ell$, the number of guanine
in the window $\sum u_i$ exactly obeys the binomial distribution
$B(\ell,p)$ \cite{Li98}. Thus the PDF of the density $f_\ell$ with
a binomial shape is asymmetrical when the  of trials is small, and
approximates to be symmetrical when the number of trials large
enough. Furthermore, if the ``success" probability $p$ of each
trial is fixed between 0 and 1 (here $p=0.25$), binomial
distribution will approximate to Gaussian distribution. So for the
larger window size, PDFs of $f_\ell$ will have a Gaussian shape.
We find that at a small scale (less than several hundred bp) the
right wing of $f_\ell$ extends further more than that of left
wing. When scale $\ell \approx 10^3$ bp, the shape of PDFs becomes
nearly symmetrical and Gaussian-like (data not shown here) which
indicates the vanishing of window size effects. We hereafter
analyze natural DNA sequences beyond this scale. Careful
calculation of PDFs make it possible to investigate the
fluctuation structures at very large scales up to $10^5$ bp, about
$1/2$ to $1/100$ of a typical microbial genome. For eukaryotic
genome such as the human chromosomes, larger scales are more
practical to study. But for comparison, the scale ranges are fixed
between $10^3$ and $10^5$ for all sequences studied below.

The evolution of the PDFs of guanine density fluctuation within
the scale range $\ell=2^{10} \sim 2^{17}$ bp for the full length
of the four sequences in Fig.~\ref{fig:fluc} are shown in
Fig.~\ref{fig:pdf} where only two scales $\ell_{min}=2^{10}$ and
$\ell_{max}=2^{17}$ are displayed. With scale $\ell$ increasing,
the distribution of tails progresses further, indicating the
emergence of highly intense fluctuations in $f_\ell$. The four
sets of PDFs show distinct shapes. Both random and simulated
sequences have the narrow shapes of PDFs, corresponding to their
low fluctuation magnitudes. Moreover, the PDFs of simulated
sequence are also symmetrical as those of random sequence. Other
two (Ecoli and Hsap4) are not symmetrical. The right wings of the
PDFs of Hsap4 are far more higher than the left with an
exponential decaying tail, and those of Ecoli is rightly opposite.
Such tendencies in the changes of PDFs can be well captured by our
quantitative HS analysis below.

\section{\label{sec:hs} Measurements}

\subsection{Extended self-similarity analysis}

\begin{figure}
\includegraphics[width=9cm,height=11cm]{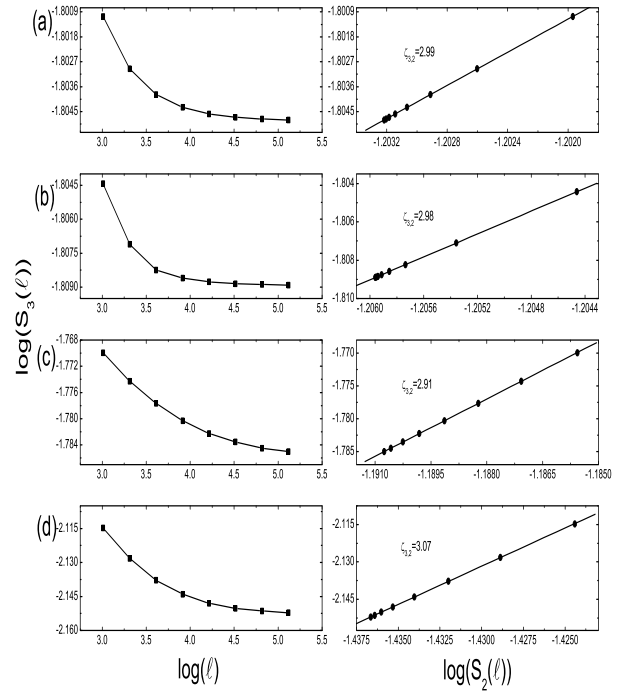}
\caption {\label{fig:sp-ess} SS and ESS plots with $S_3(\ell)$ vs.
$\ell$ and $S_3(\ell)$ vs. $S_2(\ell)$ respectively for (a)
Random, (b) Simulation, (c) Ecoli and (d) Hsap4. The scale range
is $\ell=2^{10}\sim2^{17}$. Note the curve of SS plots means no
absolute scaling, whereas the linearity of ESS plots means the
validation of relative scaling property. The ESS scaling exponents
are measured by a least square fitting.}
\end{figure}

Denote by $S_p(\ell)$ the p$th$ order moment of the fluctuation
$f_\ell$:
\begin{equation}
S_p(\ell)=\langle f_\ell^{p}\rangle = \int f_\ell^{p} P(f_\ell) d f_\ell, %
\end{equation}
where $P(f_\ell)$ is the PDF of $f_\ell$. For the calculation of
PDF, we take the linear sequence as a circle, thus all bases in
the sequence are treated equally (especially for large $\ell$). In
fact, most prokaryotic genomes are indeed circular. For large
linear eukaryotic chromosomes like those of \textit{H. sapiens}
($L\gg\ell$), the choice of open or close boundary conditions
gives essentially the same result.

In previous studies, great efforts were given to explore the power
law scaling properties of some quantities like $S_p(\ell)$ with
respect to the length sale $\ell$: $S_p(\ell)\sim \ell^{\zeta_p}$,
called self-similarity (SS), where $\zeta_p$ is called the scaling
exponents \cite{Li97}. Consequently, a log-log plot of $S_p(\ell)$
versus $\ell$ will give a straight line with a slope $\zeta_p$.
The DFA method \cite{Peng95} for analyzing ``DNA walk" is such a
SS approach. In many cases, however, such abstract scaling
property does not hold well. Therefore, a more general scaling
relation called extended self-similarity (ESS) \cite{ESS} has been
introduced in turbulence field, which is widely valid even when SS
property is not available. The existence of ESS implies that the
moments of different orders have a common changing mode with
respect to the length scales, thus the ESS is also called relative
scaling, with the form like
\begin{equation}
S_p(\ell) \sim S_q(\ell)^{\zeta_{p,q}}, %
\end{equation}
where $\zeta_{p,q}$ is called the relative scaling exponents.
log-log plots of $S_3(\ell)$ vs. $\ell$ for guanine with $\ell$
ranging from $10^3$ to $10^5$ bp are shown in
Fig.~\ref{fig:sp-ess}(a), where the bended curves indicate the
lack of power law scaling for all four sequences.
Figure~\ref{fig:sp-ess} (b) displays plots with $S_3(\ell)$ vs.
$S_2(\ell)$, where the perfect linearity verifies the existence of
the ESS property in the same scale range. Careful examines for
other $S_p(\ell)$ with higher order $p$ (up to order $p=8$) also
show the validation of ESS (data not shown here). The relationship
between scaling exponents $\zeta_{p,2}$ with different orders $p$
can be precisely predicted by the HS model as below.

\subsection{Hierarchical structure analysis}

The HS model was originally proposed by She and L\'{e}v\^{e}que
\cite{She94} to describe inertial-range multi-scale fluctuations
in terms of a similarity relation between structures of increasing
intensities of successive moment-orders $p$ in the hydrodynamics
turbulence fluid. This new similarity relation as a generalization
of the Kolmogorov's complete-scale-similarity was later developed
as a HS theory \cite{She98}, which has been successfully applied
to analyze many turbulence related field, such as the
Couette-Taylor flow \cite{She01}, flows in rapidly rotating disk
\cite{She03a}, the climate turbulence \cite{She02}, astrophysical
magnetohydrodynamic turbulence \cite{Padoan03}. and other various
complex systems, such as the diffusion-limited aggregates
\cite{QC97}, the luminosity fields of natural image
\cite{Turiel98}, chemical reaction patterns \cite{pattern}.
Preliminary analysis of the base density fluctuations at moderate
length scales along microbial genomes \cite{She01a} has given also
an encouraging sign that leads to the present work.

The HS model introduces a hierarchy of functions for successive
fluctuation intensities:
\begin{equation}
\mu_p(\ell)={S_{p+1}(\ell)\over S_p(\ell)}={{\int f_\ell^{p+1}
P(f_\ell) d f_\ell}\over {\int f_\ell^{p} P(f_\ell) d
f_\ell}}=\int f_\ell Q_p(f_\ell) d f_\ell,
\end{equation}
where $Q_p(f_\ell)={{f_\ell^{p} P(f_\ell) d f_\ell}\over {\int
f_\ell^{p} P(f_\ell) d f_\ell}}$, which is a weighted PDF for
which $\mu_p(\ell)$ is the mathematical expectation. Such a
hierarchy $\mu_p(\ell)$ covers the mean density fluctuation
intensity $\mu_0$, and a series of increasing hierarchical
intensities with increasing order $p$, and finally approaches to
the intensity of the so-called most intermittent structure,
$\mu_{\infty}(\ell)=\lim_{p\rightarrow\infty} \mu_p(\ell)$.
Therefore, one can associate each intensity with an appropriate
order $p$ which varies continuously from 0 to infinity. The
increasing hierarchical intensity reflects the increasing
contribution of positive fluctuation events while reduce that of
negative ones. When $p$ is small (less than 10), the hierarchical
function $\mu_p$ is dominated by the struggle of negative and
positive components of bold fluctuations (which is presented by
the shape of PDFs).

Note that $\mu_p(\ell)$ is a function of both $\ell$ and $p$,
which is an inherent merit of the HS model: both scales and
intensities are related to describe the multi-scaling property of
fluctuation structures. The HS model postulates a relation among
various intensities, called HS similarity, with the form like:
\begin{equation}\label{eq:hss}
\frac{\mu_{p+1}(\ell)}{\mu_{1}(\ell)}=\frac{\alpha_{p}}{\alpha_0}
\left(\frac{\mu_{p}(\ell)}{\mu_0(\ell)}\right)^\beta ,
\end{equation} %
where the exponent $\beta$ is a constant and $\alpha_p$ is
independent of $\ell$. The validation of the HS similarity
relation Eq.~(\ref{eq:hss}) can be tested by a so-called
$\beta$-test \cite{She01,She03c}, which says that a log-log plot
of $\mu_{p+1}(\ell)/\mu_{1}(\ell)$ vs. $\mu_{p}(\ell)/\mu_0(\ell)$
(often both items are normalized by the smallest scale $\ell_0$)
can be constructed, and the HS similarity is satisfied as long as
a linearity is observed. Then the HS parameter $\beta$ can be
obtained by measuring the slope using the least square fitting.
Technically speaking, this completes the HS analysis of a given
set fluctuation data.

The HS theory can construct a scaling equation to predict the ESS
scaling exponents. Equation~(\ref{eq:hss}) leads to a general
formula of the scaling exponents:
\begin{equation}
\zeta_{p,2}=\gamma p+C {(1-\beta^{p})}, %
\label{eq:zetap-general}
\end{equation}
where $C=(1-2\gamma)/(1-\beta^2)$ is determined by
$\zeta_{2,2}=1$. The parameter $\gamma$ is introduced to
characterize the most intermittent structure:
$\mu_\infty(\ell)\sim S_2^\gamma$. Note that $S_0(\ell)\equiv 1$,
$S_1(\ell)\equiv C_0$ where $C_0$ is the average density, and both
constants are independent of the scale $\ell$ and of $S_2(\ell)$,
thus we have the exact results $\zeta_{0,2}=0$, $\zeta_{1,2}=0$
and $\zeta_{1,2}=0$. The first constraint is automatically
satisfied by Eq.~(\ref{eq:zetap}), but the second constraint
introduces a relation between the parameter $\beta$ and $\gamma$
to make only one of them independent. Therefore, the HS model here
leaves only one free parameter to describe multi-scaling exponents
of the nucleotide density fluctuations. An analysis shows
\cite{Ouyang04}:
\begin{equation}
\zeta_{p,2}=\frac{(1-\beta)p-(1-\beta^p)}{(1-\beta)^2}.
\label{eq:zetap}
\end{equation}
where $\beta\neq 1$. Note the situation of $\beta\rightarrow 1$
means no intermittency, because $\gamma$ will approximate to
infinite. When $\beta=1$ the (relative) scaling exponents will be
a quadratic form $\zeta_{p,2}=p(p-1)/2$ which can be exactly
observed in a completely random DNA sequence.

The HS similarity relation Eq.~(\ref{eq:hss}) means that functions
$\mu_p(\ell)$ obey a generalized similarity relation over a range
of scales $\ell_1 \le \ell \le \ell_2$ and over a range of
intensities $p_1 \le p \le p_2$. Such a HS similarity is an
indication of the self-organization of the ensemble of the
fluctuation events. When the HS similarity is presented, the
parameter $\beta$ measures the multi-scale, multi-intensity and
self-organized property of the system \cite{She98}. When $\beta
\rightarrow 1$, the system are composed of completely self-similar
structures. The corresponding physical picture is the Kolmogorov
turbulence, where the large and small-scale statistics are
completely self-similar. We will report below if an artificial DNA
sequence is completely random, its base density fluctuations will
belong to such case. The deviation of $\beta$ from one means
intermittency. Generally speaking, More departure the $\beta$,
more outstanding the most intermittent structures stand with
respect to the background fluctuations. Therefore, the value of
$\beta$ is more intuitively related to the degree of
intermittency.

\section{\label{sec:results} HS analysis for Genomic data}

\begin{figure}
\includegraphics[width=9cm]{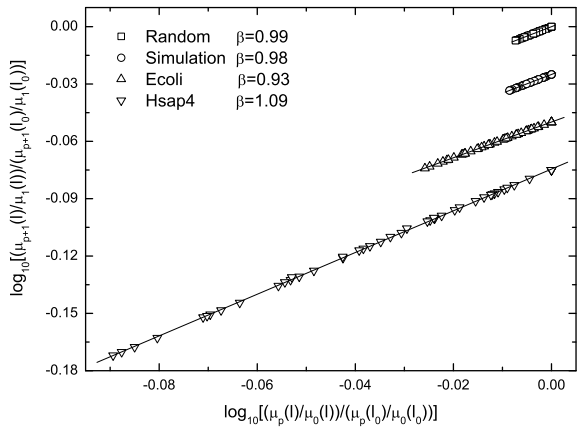}
\caption{\label{fig:beta-test} The $\beta$-test of guanine density
fluctuation for Random, Simulation, Ecoli and Hsap4 at the range
of $2^{10} \le \ell \le 2^{17}$ and $0 \le p \le 8$. A straight
line indicates the validity of the HS similarity. The slope
$\beta$ is estimated by a least square fitting. For clarity, the
second, third and fourth set of data points are displaced
vertically up by a suitable amount.}
\end{figure}

\begin{figure}
\includegraphics[width=9cm]{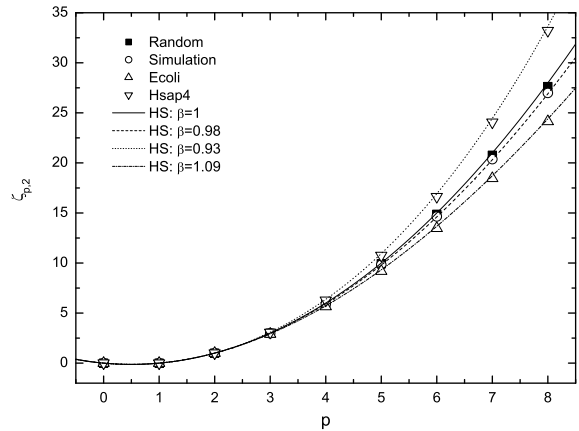}
\caption{\label{fig:zetap} ESS scaling exponents $\zeta_{p,2}$ of
guanine density fluctuation for Random, Simulation, Ecoli and
Hsap4. Dotted are the HS model formulas Eq.~(\ref{eq:zetap}) with
$\beta$ obtained from Fig.~\ref{fig:beta-test}. Solid lines
correspond to a reference $\beta=1$: $\zeta_{p,2}=p(p-1)/2$. Note
that the HS model fits exactly the four sets of scaling
exponents.}
\end{figure}

We conduct the HS analysis for variant kinds of organisms spread
all over three kingdoms of species: Eukaryote with \textit{Homo
sapiens} (24 chromosomes) and \textit{Saccharomyces cerevisiae
cerevisiae} (16 chromosomes); Prokaryote with 16 Archaea complete
genomes and 124 Bacteria complete genomes/chromosomes; and 67
viruses complete genomes publicly available in the NCBI RefSeq
Release 3, January 30, 2004. For the \textit{H. sapiens} genome,
each fully sequenced chromosome composes a few ``contigs", which
is a draft or finished sequence. Therefore, we select the longest
contig (typically large than 10 million bp) in the set of
\textit{H. sapiens} chromosomes and analyze them separately. The
chromosomes of \textit{S. cerevisiae} are completely sequenced,
and thus each was analyzed independently. The 16 Archaea genomes,
according to the Bergey's Manual of Systematic Bacteriology (2nd
edition), are belong to two phyla: 4 of \textit{Crenarchaeota} and
12 of \textit{Euryarchaeota} \cite{Bergey01}. The kingdom of
Bacteria is divided into 23 phyla by Bergey's Manual. The 124
species/strains studied here are spread over 13 phyla. Each
species/strains is assigned by their `Bergey Code' for
classification as introduced by Qi \textit{et al.} \cite{Qi04}.
For example, \textit{Escherichia coli} K12 is listed under Phylum
BXII (Proteobacteria), Class III (Gammaproteobacteria), Order XIII
(Enterobacteriales), Family I (Enterobacteriaceae), Genus XIII
(Escherichia). We change all Roman numerals to Arabic and write
the lineage as B.12.3.13.1.13. The 67 viruses sequences studied
are long enough (beyond $2^{17}$ bp) for statistic analysis.
Scales for analyzing the nucleotide fluctuations is consistent
with the above, from $10^3$ to $10^5$ bp. For each sequence, four
kinds of bases (adenine (A), cytosine (C), guanine (G) and thymine
(T)) as four fundamental ``words" in DNA sequences are analyzed
independently.

Most sequences reasonably pass the $\beta$-test (with a
correlation coefficient above 0.9995) for four different bases,
thus HS parameter $\beta$ can be obtained from the linear fitting.
The measured $\beta$ of four kinds of bases are listed in
Table~\ref{tab:virus} $\sim$ Table~\ref{tab:hsap}, for viruses,
Bacteria, Archaea, \textit{S. cerevisiae} and \textit{H. sapiens},
respectively. The $\beta$ obtained from random sequence and
simulated genome sequence with the minimal model is also listed in
Table~\ref{tab:model} to be compared. Some other important
information about these sequences, such as names, NCBI accession
numbers (Acc. No.), Bergey codes, lengths, base compositions are
also listed in the tables. Items are ordered in according to the
Bergey code, so species/strains closely related are listed
together.

\subsection{The meaning of parameter $\beta$}\label{subsec:meaning}

\begin{figure}
\includegraphics[width=9cm,height=9cm]{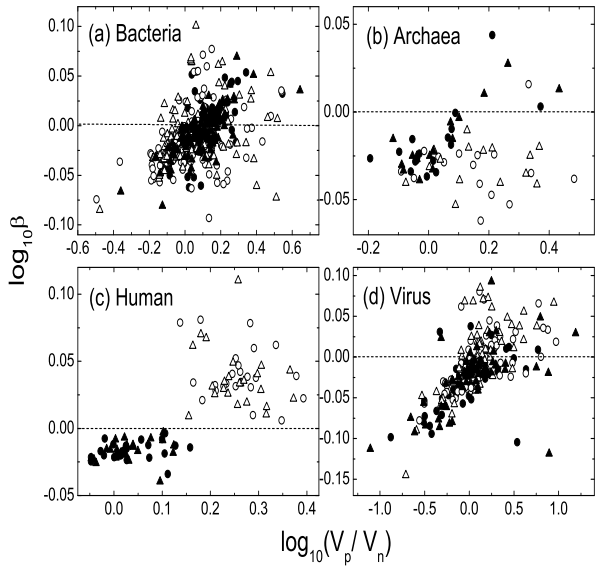}
\caption{\label{fig:flux} The log-log plot of $\beta$ vs.
$V_p/V_n$ for (a) Bacteria, (b) Archaea, (c) Human chromosomes and
(d) viruses. $V_p$ and $V_n$ is calculated in a window with 1024
bp. Note that $\beta$ roughly increases with increased amounts
$V_p/V_n$. Where symbols are: ($\bullet$) A, ($\blacktriangle$) T,
($\circ$) C, ($\vartriangle$), G. Dash lines indicate $\beta=1$. }
\end{figure}

As an example, the results of the $\beta$-test for Random,
Simulation, Ecoli and Hsap4 are shown in Fig.~\ref{fig:beta-test},
where the scale range is between $10^3$ bp and $10^5$ bp and
$\ell_0=1024$. The exactly good linearity of plots for all four
cases indicate that the HS similarity is satisfied, which means
that all genomic sequences including the random one have a nicely
self-organized HS scaling property. The values of $\beta_G$
obtained are $0.99 \pm 0.000$, $0.98 \pm 0.000$, $0.93 \pm 0.001$
and $1.09\pm 0.002$ for Random, Simulation, Ecoli, and Hsap4,
respectively. ESS relative scaling exponents $\zeta_{p,2}$
measured in Fig.~\ref{fig:sp-ess} as a function of order $p$ are
plotted in Fig.~\ref{fig:zetap}, where also presents the
prediction of HS model Eq.~(\ref{eq:zetap}) with parameters
$\beta$ obtained in Fig.~\ref{fig:beta-test}. Good agreements
between the data of scaling exponents (points) and HS model
predictions (lines) are exactly established. The result of the
random sequence analyzed shows that its scaling exponents have a
theoretical quadratic form $\zeta_{p,2}=p(p-1)/2$ with $\beta$=1.
Furthermore, scaling exponents $\zeta_{p,2}$ are distinctly
separated into three groups: Random and Simulation with
systematically moderate $\zeta_{p,2}$; Hsap4 with larger ones;
Ecoli with smaller ones. Theoretical speaking, smaller ESS scaling
exponents mean less heterogeneous and high intermittent
corresponding to the smaller $\beta$. Results of
Fig.~\ref{fig:zetap} are consistent well with this theoretical
consideration.

The quantitative $\beta$ values are different for these four
sequences. The \textit{i.i.d.} random sequence has a $\beta$ very
close to 1 which is consistent with the self-similarity picture of
both its base density fluctuations and the Gaussian-like PDFs.
Simulated sequence with $\beta$ close to 1 indicates that its base
density fluctuations are very near to random. Interestingly, Ecoli
and Hsap4 have different $\beta$ values with remarkable deviations
from one, where $\beta_G$ of Ecoli is lower than one and that of
Hsap4 is on the opposite. This contrast can also be seen from the
opposite fluctuations of the two sequences in Fig.~\ref{fig:fluc}.
The deviation of $\beta$ from one is consistent with the
increasing of more intermittent structures presented in the
guanine density fluctuations shown in Fig.~\ref{fig:fluc} and
Fig.~\ref{fig:pdf}. The $\beta$ values measured here give a well
quantitative description of this intermittent structures.

Most $\beta$ values for various genomes in Table~\ref{tab:virus}
$\sim$ Table~\ref{tab:hsap} significantly deviate from one,
indicating a non-Gaussian statistical property of the base density
fluctuations. Mathematically, the existence of $\beta$ in a range
$\ell_1<\ell<\ell_2$ means that the incremental hierarchical
intensity $\mu_p$ (p=0,1,...) in this scaling range are linked by
a similarity parameter $\beta$. $\mu_p$ is the mathematical
expectation of $p$ order weighted PDF which is directly related to
the shape of the original PDF (0-order). If the original PDF is
skewed or peaked, the incremental rate of $\mu_p$ will deviate
from that of a Gaussian PDF, which will lead to the deviation of
$\beta$ from one. Thus $\beta$ is closely related to the
heterogeneity (caused by atypical base density) of the sequence.
The measured $\beta$, that is dependent on individual organisms,
reflects different fluctuation structures of the different
genomes.

For illustrating this point, we study the atypical components of
the fluctuation signals. When atypical components are biased
distribution, e.g., with more positive components than negative
ones, the PDF is skewed with a long right tail. $V_p$ is
introduced to measure the percentage of the positive components
beyond a threshold relative to the whole ensemble:
\begin{equation}
V_p = \int_{\mu+H}^{1} f P(f) d f,
\end{equation}
where $f$ is the base composition measured in a fixed window and
$\mu$ is the mean value of $f$. Similarly, the percentage of the
negative components $V_n$ is defined as
\begin{equation}
V_n = \int_{0}^{\mu-H} f P(f) d f.
\end{equation}
When calculating $V_p$ and $V_n$, we fix the window length to be
1024 bp and let $H$ be 1.5 times of standard deviation of $f$. For
reasonable PDF (with a single maximum), the skewness can be
roughly characterized by the relative value of $V_p$ and $V_n$.
When $V_p$ is far more than $V_n$, the PDF tends to have a long
right tail (such as Fig.~\ref{fig:pdf}(d)), and vice versa. We
study the relationship between $\beta$ and the biased fluctuation
of local base density by a log-log plot $\beta$ vs. $V_p/V_n$ in
Fig.~\ref{fig:flux} for Bacteria, Archaea, Human chromosomes and
viruses respectively. Note that $\beta$ and $V_p/V_n$ have the
same tendency, which relate $\beta$ with the biased distribution
of atypical components. In a word, $\beta$ measure the
heterogeneity of a genome sequence. Although a theory called
mutational equilibrium theory \cite{Sueoka62} for the
interpretation of the stationarity of G+C content within a species
has been proposed, the understanding (both qualitative and
quantitative) on base compositional heterogeneity is still
limited. Hereafter we propose $\beta$ as the ``prob" to
systematically study the genomic heterogeneity.

The analysis above is focused on the meaning of parameter $\beta$
in terms of fluctuations of base composition, which is a
intuitional study on the physical picture of genome sequences.
Such a physical study may have other biological implications. It
should be emphasized that HS theory has an mathematical invariance
(symmetry) by defining a transformation group \cite{She98}. Such a
symmetry is exactly achieved through a log-Poisson cascade process
\cite{poisson}. When we carry out multiscaling and hierarchical
analysis of DNA, RNA and protein sequences, some transformations
or symmetry may be useful to clarify the complexity of a
biological system, especially genome data which combine the
information of the structure and function together. We think that
these quantitative properties, especially HS parameter $\beta$,
are revelatory to characterize biological questions and farther
research should be done. In the following, we will expound two
points respectively: DNA strand symmetry in
subsec.~\ref{subsec:symmetry} and sequence complexity with $\beta$
in subsec.~\ref{subsec:compl}.

\subsection{\label{subsec:symmetry} Strand symmetry}

\begin{figure}
\includegraphics[width=9cm]{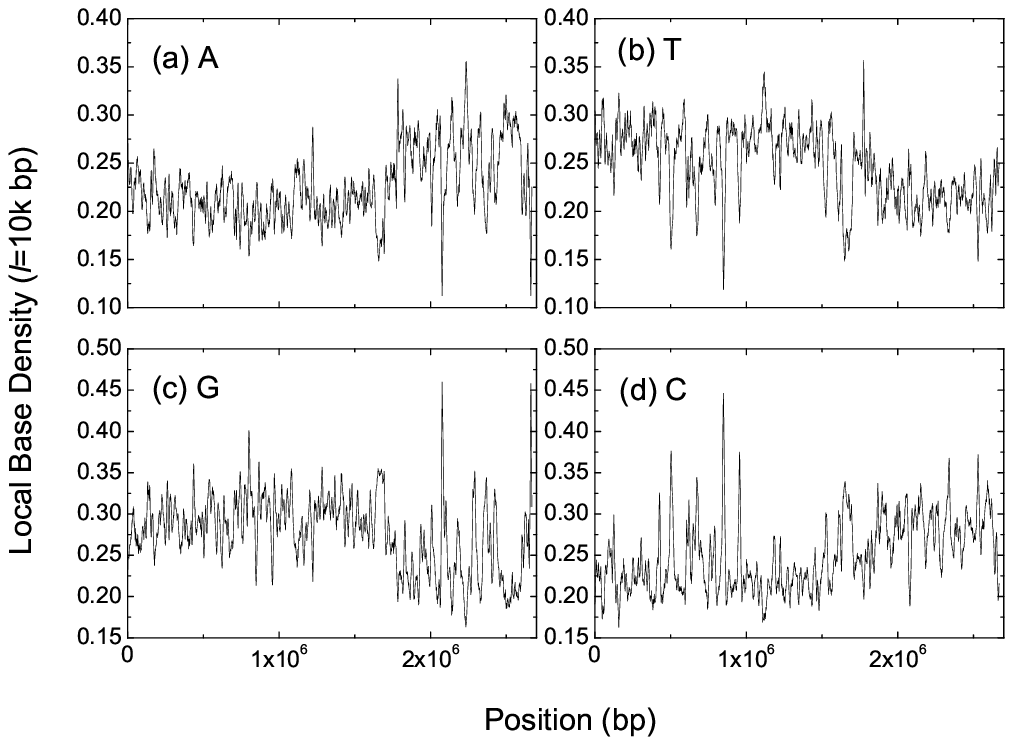}
\includegraphics[width=9cm]{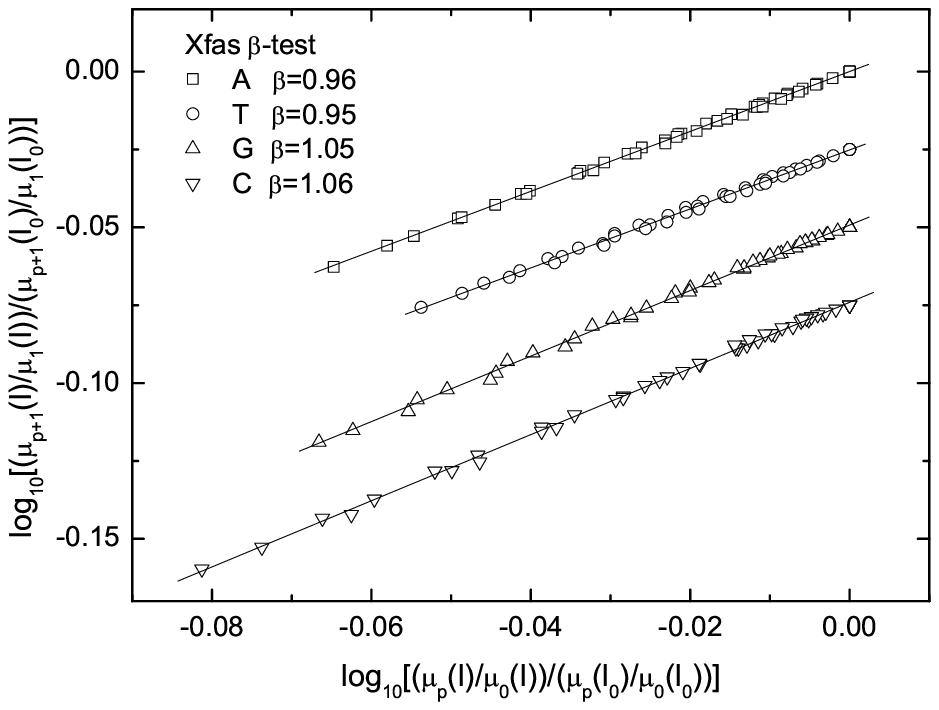}
\caption {\label{fig:parity} (Top) Local density fluctuations with
a scaling window of $\ell=10^4$ bp of four bases (a) A, (b) T, (c)
G and (d) C for \textit{B-Xfa}. The sliding window moves at a step
of length $\Delta=10^3$ bp. Note the approximately enantiomorphous
fluctuation between A and T, G and C; complementary property of
base fluctuation between A and G, T and C. (Bottom) $\beta$-test
of base density fluctuation for \textit{B-Xfa} at the range of
$2^{10} \le \ell \le 2^{17}$ and $0 \le p \le 8$. A straight line
indicates the validity of the HS similarity. Note that the parity
rule $\beta_A\approx\beta_T$ and $\beta_C\approx\beta_G$ is
obeyed. For clarity, the second, third and fourth set of data
points are displaced vertically up by a suitable amount.}
\end{figure}

\begin{figure}
\includegraphics[width=9cm]{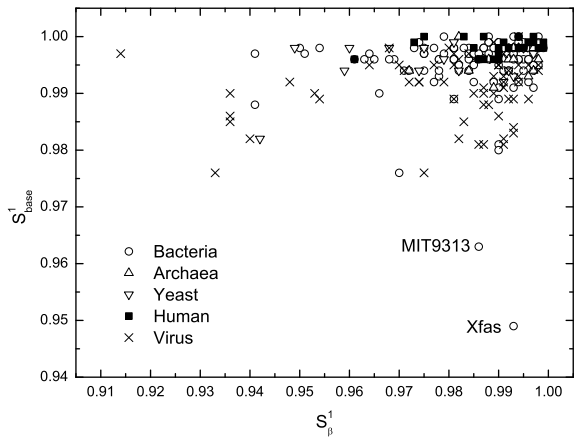}
\caption{\label{fig:symmetry_beta} The symmetry levels
$S^{1}_{\beta}$ measured on the base $\beta$ values of genomes or
chromosomes are increased with the increasing sequence length.}
\end{figure}

One of the most intriguing results obtained here is that base
composition fluctuations of most Prokaryotic genomes and
Eukaryotic chromosomes obey a parity rule: $\beta_A\approx\beta_T$
and $\beta_C\approx\beta_G$. This is the extension of Chargaff's
parity rule 2 (PR2), which states that if \textit{single strands}
of a long DNA duplex (say, a few thousand bp) are isolated and
their base compositions are determined, then $\%A \cong \%T$, and
$\%C \cong \%G$ \cite{Forsdyke02}. The validity of PR2 became
clearer when full genome sequences are calculated. PR2 can be
generalized to compositions of dinucleotide and other
oligonucleotide \cite{Prabhu93}, or even the whole base-base
correlation function \cite{Li97,Teitelman96}. PR2 is generally
interpreted as the strand symmetry of biological functionalities
such as mutation and/or selection pressures acting on single base
or oligonucleotide. Local asymmetrical base composition is also
numerously reported \cite{Mrazek98}. As illustrated in
Fig~\ref{fig:parity} (Top), the local base density of A(G) and
T(C) is not equal, but enantiomorphous each other. However, we
find the $\beta$ values of each pair are very close. The parity
rule in terms of fluctuation discovered here means that when full
genomic sequences are considered, fluctuation structures of A(C)
and T(G) are approximately identical, although the local
fluctuation at the same position may be different. This point is
well illustrated in Fig.~\ref{fig:parity} (Bottom) in spit of
remarkable out-of-phase fluctuations. The existence of PR2 in
terms of HS parameter $\beta$ means that the global function of
evolutionary factors, such as mutation pressure or selection
pressure, are not bias on a genomic scale. This finding is
consistent with our previous results \cite{Ouyang04}, and such a
biological implication may be deserved to study in the future.
Recently ref.~\cite{Baisnee02} introduced a similarity function to
describe the strand symmetry, which has the form like:
\begin{equation}
S^1=1-{{|f_{A}-f_{T}|+|f_{C}-f_{G}|}\over
{|f_{A}+f_{T}|+|f_{C}+f_{G}|}},
\end{equation}
where $f_i$ with $i=\{A,T,C,G\}$ denote the density of any single
nucleotide. $S^1$ can be used to characterize the symmetry level
with a range from 0 (asymmetry/dissimilarity) to 1 (prefect
symmetry/similarity) (details can be found in
Ref.~\cite{Baisnee02}). We calculate $S_{\beta}^{1}$ on the
$\beta$ of four bases, and display the results in
Fig.~\ref{fig:symmetry_beta}, where the symmetry levels roughly
increase with the increasing of sequence length.

\subsection{\label{subsec:compl} Sequence complexity}

\begin{figure}
\includegraphics[width=9cm]{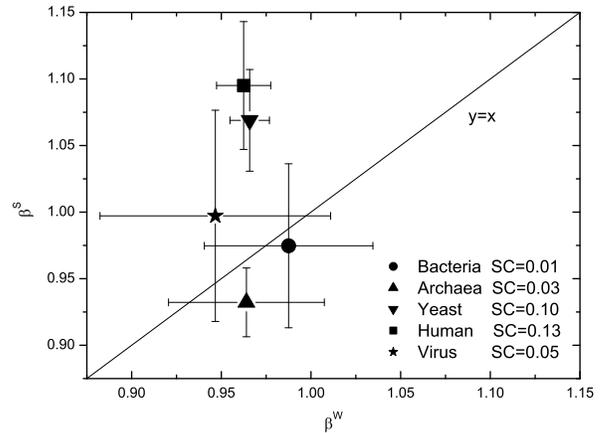}
\caption{\label{fig:beta_cluster} The mean values of
($\beta_C$+$\beta_G$)/2 versus ($\beta_A$+$\beta_T$)/2 for 124
bacterial genomes/chromosomes, 16 archaeal genomes, Yeast, Human
and Viruse. The solid points are the mean values  Note the cluster
property of the three kingdoms and the diversity of viruses.}
\end{figure}

Another intriguing results is the systematic change of $\beta$
with evolutionary categories. The relationship between
evolutionary categories and sequences correlation structures have
been studied previously in \cite{Voss92,Peng95,compl}. Note that
Buldyrev \textit{et al.} \cite{Buldyrev95} suggested that the
complexity of noncoding DNA sequences increased with evolution,
with an increasing of spectrum exponents for highly evolved
species. While Voss \cite{Voss92} found the spectrum exponents
decrease with evolution. These incompatible findings are due to
the equivocal meaning of spectrum exponents. We have shown that HS
parameter $\beta$ has an implication of biological evolution
\cite{Ouyang04}, that is, the decrease of category averaged
$\beta_A$ reflects the increasing degree of organization in more
developed species. As shown in Sec.~\ref{subsec:meaning}, we
related the decrease of $\beta$ (of a specific base) with the
increasing sequence heterogeneity introduced by concentration of
low-density base compositions.

We introduce a new definition of sequence complexity as the total
heterogeneity of the four different bases. Because of strand
symmetry, we reduce the number of variables from four to two by
setting $\beta^{S}=(\beta_C+\beta_G)/2$ and
$\beta^{W}=(\beta_A+\beta_T)/2$. Then the quantitative expression
for sequence complexity is written as:
\begin{equation}
SC=|\beta^{S}-\beta^{W}|.
\end{equation}
For a random sequence, $SC=0$ because in that case $\beta_A
\approx \beta_G$. It establish a zero complexity for random
sequences. Complexity of simulated sequence generated by the
minimal model \cite{Hsieh03} is nearly zero. The SC values for
different categories, which is shown in
Fig.~\ref{fig:beta_cluster}, indicate that Human has the highest
complexity and Eukaryotes has a higher complexity than
Prokaryotes. Interestingly, on average Archaea is more complex
than Bacteria. It is not clear if this is because the number of
bacterial genomes studied is sufficient to get a statistical
average while that of archaeal genomes is biased by its relatively
small number of members. But it is remarkable that in this finite
set both $\beta^{S}$ and $\beta^{W}$ of archaea genomes are lower
than most of bacteria genomes. This is an interesting phenomenon
which needs further study. Virus are very diverse: their $\beta$
values have a big variance. This may be related to their
variability nature.

The relative magnitude of $\beta^{S}$ and $\beta^{W}$ is also
needs further investigation. From Fig.~\ref{fig:beta_cluster} it
is clear that for Human $\beta_{S}>\beta_{W}$, while Archaea and
Bacteria are on the contrary. It indicates the presence of many
low-concentration regions of C or G (and hence a high
concentration of A or T) in the genomes of prokaryotic genomes.
While many regions with high concentration of C or G are spread
along the Human genome.

\begin{figure}
\includegraphics[width=9cm]{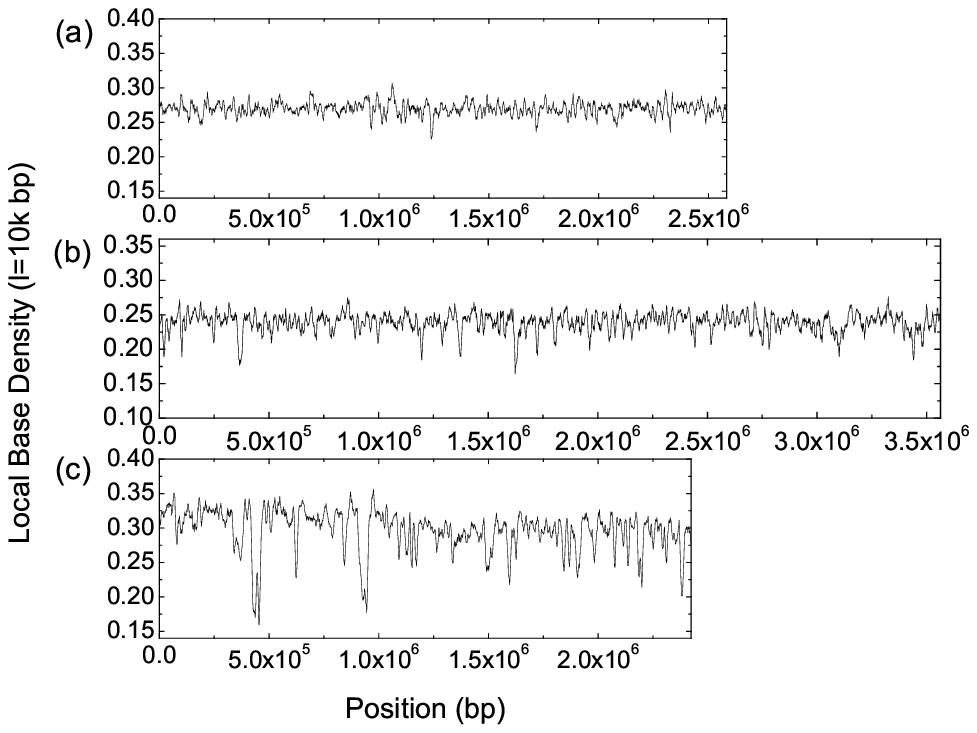}
\includegraphics[width=9cm]{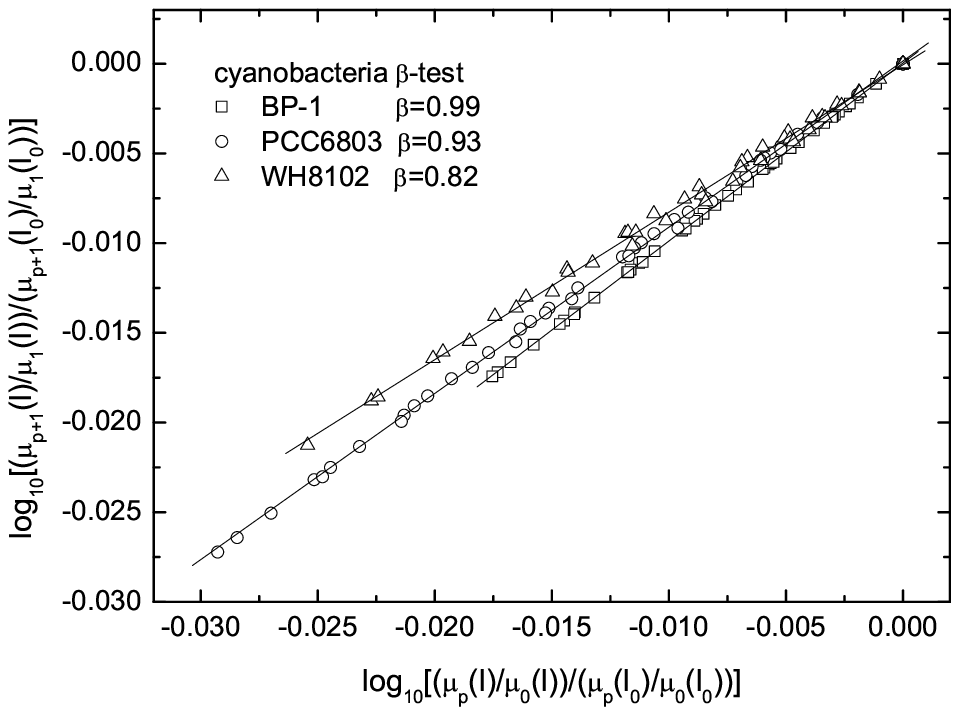}
\caption {\label{fig:syne} (Top) Local density fluctuations with a
scaling window of $\ell=10^4$ bp of G for (a) BP-1, (b) PCC6803
and (c) WH8102. The sliding window moves at a step of length
$\Delta=10^3$ bp. (Bottom) $\beta$-test for BP-1, PCC6803 and
WH8102. The test range is $2^{10} \le \ell \le 2^{17}$ and $0 \le
p \le 8$. A straight line indicates the validity of the HS
similarity. Note the smaller value of $\beta$ for WH8102
corresponds to many low-concentration regions of G.}
\end{figure}

One plausible origin of sequence complexity is horizontal gene
transfer (HGT), which has been recognized as one of the major
forces in prokaryotic genome evolution \cite{Koonin01}. HGT
increase the heterogeneity by incorporating alien sequences,
because recent transferred sequences from distantly related
species have not undergone sufficient mutational pressure, thus
its atypical base composition can be distinguished from ancestral
DNA \cite{HGT,Garcia00}. According to this criteria, Garcia-Vallve
\textit{et. al.} \cite{Garcia00} found that 0$\%$ to 22.2$\%$ of
total genes of 88 bacterial and archaeal genomes are obtained by
horizontal gene transfer. The guanine density fluctuation  and
$\beta$-test of three typical cyanobacteria (Thermosynechococcus
elongatus BP-1 (BP-1), Synechocystis PCC6803 (PCC6803) and
Synechococcus sp. WH8102 (WH8102)) are shown in
Fig.~\ref{fig:syne}. WH8102 has a notable small $\beta$ and
extensive low-guanine regions comparing to the other two. Indeed,
a lot of low G+C segments of the genome of WH8102 have been
comprehensively identified as obtained by HGT \cite{Palenik03},
contributing to its functionalization of the envelope-modification
of the cell surface and the motility of swimming.

\section{\label{sec:con} Conclusions}

Our approach by HS analysis have some merits as the following:
first, it has a solid theoretical foundation and has obtained many
experimental supports, and varies models deduced from the HS
theory has been widely used to analyze nonlinear fields containing
intermittent structures; second, It employs an extended scaling
analysis which lead to more accurate identification of scaling
property; third but not least, the multiple scale fluctuation
analysis of base composition is adequate to detect large scale (up
to $10^5$ bp) correlations.

By carrying out an systematic study of the large scale structures
of available genomes, we show that the large scale base density
fluctuation ($10^3-10^5$ bp) of most examined sequences (including
genomes of Archaea, Bacteria, Eukaryotes and viruses) satisfy the
HS similarity relation. It reveals that base density fluctuations
of genomes are hierarchically organized across scales and across
intensities. The HS analysis ($\beta$-test) allows one to quantify
the degree of multi-scale and multi-intensity correlations. It is
known that the major contributing factors to the sequence-wide
pattern is not the mean base content and correlations among
neighboring bases in genome sequence, but the spatial
heterogeneity of the base composition fluctuation or the
long-range correlation that largely shapes the complexity of the
whole sequence. Our HS parameter $\beta$ obtained can capture this
point. Furthermore, $\beta$ is effective to describe horizontal
gene transfer, strand symmetry, phylogenetic relations of various
biological species.

It is shown the values of $\beta$ for natural DNA sequences show
distinct deviation from one which is illustrated as the case of a
completely random one. $\beta$ values show significant divergence,
but preserve the parity among bases presumably the consequence of
strand symmetry. The HS parameter $\beta$ are clustered according
to evolution categories. It indicate that spatial heterogeneity of
base composition or long-range correlation in genomic sequences
are different for Archaea, Bacteria and Eukaryotes. The
heterogeneity is interpreted as different genetic material
transfer patterns for evolutionary communities.

\acknowledgments

We have benefited from useful discussions with many people at the
LTCS and CTB of Peking University, especially Dr. Huaiqiu Zhu.
This work was supported by NNSFC No. 10225210 and by 973 Project
Grant No. 2003CB715905 founded by MOSTof China.

\appendix
\section{Abbreviations}
The abbreviations of the sequence name are alphabetically listed
in the parenthesis:

(a) Table~\ref{tab:virus}: Virus (Most abbreviations are taken
form http://www.ncbi.nlm.nih.gov/ICTVdb): African swine fever
virus (\textit{V-AsFV}), Agrotis segetum granulovirus
(\textit{V-AsGV}), Amsacta moorei entomopoxvirus
(\textit{V-AmEPV}), Autographa californica nucleopolyhedrovirus
(\textit{V-AcNPV}), Bovine herpesvirus 1 (\textit{V-BoHV-1}),
Bovine herpesvirus 5 (\textit{V-BoHV-5}), Bovine papular
stomatitis virus (\textit{V-BPSV}), Callitrichine herpesvirus 3
(\textit{V-CalHV-3}), Camelpox virus (\textit{V-CMLV}), Canarypox
virus (\textit{V-CNPV}), Cercopithecine herpesvirus 1
(\textit{V-CeHV-1}), Chimpanzee cytomegalovirus (\textit{V-CCMV}),
Choristoneura fumiferana defective nucleopolyhedrovirus
(\textit{V-CfDEFNPV}), Cowpox virus (\textit{V-CPXV}), Ectocarpus
siliculosus virus (\textit{V-EsV}), Ectromelia virus
(\textit{V-ECTV}), Equine herpesvirus 1 (\textit{V-EHV-1}), Equine
herpesvirus 2 (\textit{V-EHV-2}), Equine herpesvirus 4
(\textit{V-EHV-4}), Fowlpox virus (\textit{V-FWPV}), Gallid
herpesvirus 2 (\textit{V-GaHV-2}), Gallid herpesvirus 3
(\textit{V-GaHV-3}), Goatpox virus (\textit{V-GTPV}), Helicoverpa
armigera nucleopolyhedrovirus G4 (\textit{V-HaNPV}), Heliothis zea
virus 1 (\textit{V-HzV-1}), Human herpesvirus 1
(\textit{V-HHV-1}), Human herpesvirus 2 (\textit{V-HHV-2}), Human
herpesvirus 4 (\textit{V-HHV-4}), Human herpesvirus 5
(\textit{V-HHV-5}), Human herpesvirus 6 (\textit{V-HHV-6}), Human
herpesvirus 6B (\textit{V-HHV-6B}), Human herpesvirus 7
(\textit{V-HHV-7}), Human herpesvirus 8 (\textit{V-HHV-8}),
Ictalurid herpesvirus 1 (\textit{V-IcHV-1}), Invertebrate
iridescent virus 6 (\textit{V-IIV-6}), Lumpy skin disease virus
(\textit{V-LSDV}), Lymantria dispar nucleopolyhedrovirus
(\textit{V-LdNPV}), Lymphocystis disease virus - isolate China
(\textit{V-LCDV}), Macaca mulatta rhadinovirus (\textit{V-MMRV}),
Mamestra configurata nucleopolyhedrovirus A
(\textit{V-MacoNPV-A}), Mamestra configurata nucleopolyhedrovirus
B (\textit{V-MacoNPV-B}), Melanoplus sanguinipes entomopoxvirus
(\textit{V-MsEPV}), Meleagrid herpesvirus 1 (\textit{V-MeHV-1}),
Molluscum contagiosum virus (\textit{V-MCV}), Monkeypox virus
(\textit{V-MPXV}), Mouse cytomegalovirus 1 (\textit{V-MCMV-1}),
Myxoma virus (\textit{V-MYXV}), Orf virus (\textit{V-ORFV}),
Orgyia pseudotsugata multicapsid nucleopolyhedrovirus
(\textit{V-OpMNPV}), Ostreid herpesvirus 1 (\textit{V-OsHV-1}),
Paramecium bursaria Chlorella virus 1 (\textit{V-PBCV-1}),
Psittacid herpesvirus 1 (\textit{V-PsHV-1}), Rabbit fibroma virus
(\textit{V-SFV}), Rabbitpox virus (\textit{V-RPXV}), Rachiplusia
ou multiple nucleopolyhedrovirus (\textit{V-RoMNPV}), Rat
cytomegalovirus (\textit{V-RCMV}), Sheeppox virus
(\textit{V-SPPV}), Shrimp white spot syndrome virus
(\textit{V-SWSSV}), Spodoptera exigua nucleopolyhedrovirus
(\textit{V-SpeiNPV}), Spodoptera litura nucleopolyhedrovirus
(\textit{V-SpltNPV}), Swinepox virus (\textit{V-SWPV}), Tupaia
herpesvirus (\textit{V-TuHV}), Vaccinia virus (\textit{V-VACV}),
Variola virus (\textit{V-VARV}), Xestia c-nigrum granulovirus
(\textit{V-XecnGV}), Yaba monkey tumor virus (\textit{V-YMTV}),
Yaba-like disease virus (\textit{V-YLDV}).

(b) Table~\ref{tab:bacteria}: Bacteria: \textit{Agrobacterium
tumefaciens} strain C58 C \& L (\textit{B-Atu1} \&
\textit{B-Atu2}), \textit{Aquifex aeolicus} (\textit{B-Aae}),
\textit{Bacillus anthracis} A2012 (\textit{B-Ban}),
\textit{Bacillus halodurans} (\textit{B-Bha}), \textit{Bacillus
subtilis} (\textit{B-Bsu}), \textit{Bacteroides thetaiotaomicron}
VPI-5482 (\textit{B-Bth}), \textit{Bifidobacterium longum} NCC2705
(\textit{B-Blo}), \textit{Bordetella bronchiseptica}
(\textit{B-Bbr}), \textit{Bordetella parapertussis}
(\textit{B-Bpa}), \textit{Bordetella pertussis} (\textit{B-Bpe}),
\textit{Borrelia burgdorferi} (\textit{B-Bbu}),
\textit{Bradyrhizobium japonicum} (\textit{B-Bja}),
\textit{Brucella melitensis} chromosome I \& II (\textit{B-Bme1}
\textit{Brucella suis} chromosome I \& II (\textit{B-Bsu1} \&
\textit{B-Bsu2}), \& \textit{B-Bme2}), \textit{Buchnera
aphidicola} (\textit{B-Bap}), \textit{Buchnera aphidicola} Sg
(\textit{B-BapS}), \textit{Buchnera} sp. APS (\textit{B-Bsp}),
\textit{Campylobacter jejuni} (\textit{B-Cje}),
\textit{Caulobacter crescentus} (\textit{B-Cre}),
\textit{Chlamydia muridarum} (B-\textit{Cmu}), \textit{Chlamydia
trachomatis} (\textit{B-Ctr}), \textit{Chlamydophila caviae} GPIC
(\textit{B-Cca}), \textit{Chlamydophila pneumoniae} AR39, CWL029,
J138 \& TW-183 (\textit{B-Cpn1}, \textit{B-Cpn2}, \textit{B-Cpn3}
\& \textit{B-Cpn4}), \textit{Chlorobium tepidum} TLS
(\textit{B-Cte}), \textit{Chromobacterium violaceum} ATCC 12472
(\textit{B-Cvi}), \textit{Clostridium acetobutylicum} ATCC824
(\textit{B-Cac}), \textit{Clostridium perfringens}
(\textit{B-Cpe}), \textit{Clostridium tetani} E88
(\textit{B-Cte}), \textit{Corynebacterium efficiens} YS-314
(\textit{B-Cef}), \textit{Corynebacterium glutamicum}
(\textit{B-Cgl}), \textit{Coxiella burnetii} (\textit{B-Cbu}),
\textit{Deinococcus radiodurans} chromosome 1 \& 2
(\textit{B-Dra1} \& \textit{B-Dra2}), \textit{Escherichia coli}
CFT073, K12, O157:H7 \& O157:H7 EDL933 (\textit{B-Eco1},
\textit{B-Eco2}, \textit{B-Eco3} \& \textit{B-Eco4}),
\textit{Fusobacterium nucleatum} ATCC 25586 (\textit{B-Fnu}),
\textit{Haemophilus ducreyi} 35000HP (B-\textit{Hdu}),
\textit{Haemophilus influenzae} Rd (\textit{B-Hin}),
\textit{Helicobacter hepaticus} (\textit{B-Hhe}),
\textit{Helicobacter pylori} 26695 \& J99 (\textit{B-Hpy1} \&
\textit{B-Hpy2}), \textit{Lactobacillus plantarum}
(\textit{B-Lpl}), \textit{Lactococcus lactis} sp. IL1403
(\textit{B-Lla}), \textit{Leptospira intrerrogans} I \& II
(\textit{B-Lin1} \& \textit{B-Lin2}), \textit{Listeria innocua}
(\textit{B-Lin}), \textit{Listeria monocytogenes} EGD-e
(\textit{B-Lmo}), \textit{Mesorhizobium loti} (\textit{B-Mlo}),
\textit{Mycobacterium bovis subsp. bovis} AF2122/97
(\textit{B-Mbo}), \textit{Mycobacterium leprae} TN
(\textit{B-Mle}), \textit{Mycobacterium tuberculosis} CDC1551 \&
H37Rv (\textit{B-Mtu1} \& \textit{B-Mtu2}), \textit{Mycoplasma
gallisepticum} (\textit{B-Mga}), \textit{Mycoplasma genitalium}
(\textit{B-Mge}), \textit{Mycoplasma penetrans} (\textit{B-Mpe}),
\textit{Mycoplasma pneumoniae} (\textit{B-Mpn}),
\textit{Mycoplasma pulmonis} UAB CTIP (\textit{B-Mpu}),
\textit{Neisseria meningitidis} MC58 \& Z2491 (\textit{B-Nme1} \&
\textit{B-Nme2}), \textit{Nostoc} sp. PCC7120 (\textit{B-Nsp}),
\textit{Oceanobacillus iheyensis} (\textit{B-Oih}),
\textit{Pasteurella multocida} PM70 (\textit{B-Pmu}),
\textit{Pirellula} sp. (\textit{B-Psp}), \textit{Porphyromonas
gingivalis} W83 (\textit{B-Pgi}), \textit{Prochlorococcus marinus}
CCMP1375, CCMP1378 \& MIT9313 (\textit{B-Pma1}, \textit{B-Pma2} \&
\textit{B-Pma3}), \textit{Pseudomonas aeruginosa} PA01
(\textit{B-Pae}), \textit{Pseudomonas putida} KT2440
(\textit{B-Ppu}), \textit{Pseudomonas syringae} (\textit{B-Psy}),
\textit{Rickettsia conorii} (\textit{B-Rco}), \textit{Rickettsia
prowazekii} (\textit{B-Rpr}), \textit{Ralstonia solanacearum}
chromosome (\textit{B-Rso}), \textit{Salmonella typhi}
(\textit{B-Sty1}), \textit{Salmonella typhimurium} LT2
(\textit{B-Sty2}), \textit{Salmonella typhi} y2 (\textit{B-Sty3}),
\textit{Sinorhizobium meliloti} 1021, pSymA \& pSymB
(\textit{B-Sme1}, \textit{B-Sme2} \& \textit{B-Sme3}),
\textit{Shewanella oneidensis} MR-1 (\textit{B-Son}),
\textit{Shigella flexneri} 2a strain 301 (\textit{B-Sfl}),
\textit{Staphylococcus aureus} Mu50, MW2 \& N315 (\textit{B-Sau1},
\textit{B-Sau1} \& \textit{B-Sau1}), \textit{Staphylococcus
epidermidis} ATCC 12228 (\textit{B-Sep}), \textit{Streptococcus
agalactiae} 2603 V/R \& NEM316 (\textit{B-Sag1} \&
\textit{B-Sag2}), \textit{Streptococcus mutans} UA159
(\textit{B-Smu}), \textit{Streptococcus pneumoniae} R6 \& TIGR4
(\textit{B-Spn1} \& \textit{B-Spn2}), \textit{Streptococcus
pyogenes} MGAS8232, MGAS315, SF370 \& SSI-1 (\textit{B-Spy1},
\textit{B-Spy2}, \textit{B-Spy3} \& \textit{B-Spy4}),
\textit{Streptomyces coelicolor} A3(2) (\textit{B-Sco}),
\textit{Streptomyces avermitilis} MA-4680 (\textit{B-Sav}),
\textit{Synechococcus} sp. WH8102 (\textit{B-SspW}),
\textit{Synechocystis} sp. PCC6803 (\textit{B-SspP}),
\textit{Treponema pallidum} (\textit{B-Tpa}),
\textit{Thermoanaerobacter tengcongensis} (\textit{B-Tte}),
\textit{Thermosynechococcus elongatus} BP-1 (\textit{B-Tel}),
\textit{Thermotoga maritima} (\textit{B-Tma}), \textit{Ureaplasma
urealyticum} (\textit{B-Uur}), \textit{Vibrio cholerae} chromosome
1 \& 2 (\textit{B-Vch1} \& \textit{B-Vch2}), \textit{Vibrio
parahaemolyticus} RIMD 2210633 chromosome 1 \& 2 (\textit{B-Vpa1}
\& \textit{B-Vpa2}), \textit{Vibrio vulnificus} CMCP6 chromosome I
\& II (\textit{B-Vvu1} \& \textit{B-Vvu2}), \textit{Wigglesworthia
brevipalpis} (\textit{B-Wbr}), \textit{Wolinella succinogenes}
(\textit{B-Wsu}), \textit{Xanthomonas axonopodis citri} 306
(\textit{B-Xax}), \textit{Xanthomonas campestris} ATCC 33913
(\textit{B-Xca}), \textit{Xylella fastidiosa} (\textit{B-Xfa}),
\textit{Yersinia pestis} strain C092 \& KIM (\textit{B-Ype1} \&
\textit{B-Ype2}).

(c) Table~\ref{tab:archaea}: Archaea: \textit{Aeropyrum pernix}
(\textit{A-Ape}), \textit{Archaeoglobus fulgidus}
(\textit{A-Afu}), \textit{Halobacterium} sp. NRC-1
(\textit{A-Hsp}), \textit{Methanobacterium thermoautotrophicum}
(\textit{A-Mth}), \textit{Methanococcus jannaschii}
(\textit{A-Mja}), \textit{Methanopyrus kandleri} AV19
(\textit{A-Mka}), \textit{Methanosarcina acetivorans}
(\textit{A-Mac}), \textit{Methanosarcina mazei} Goe1
(\textit{A-Mma}), \textit{Pyrococcus abyssi} (\textit{A-Pab}),
\textit{Pyrococcus furiosus} (\textit{A-Pfu}), \textit{Pyrococcus
horikoshii} (\textit{A-Pho}), \textit{Sulfolobus solfataricus}
(\textit{A-Sso}), \textit{Sulfolobus tokodaii} (\textit{A-Sto}),
\textit{Thermoplasma acidophilum} (\textit{A-Tac}),
\textit{Thermoplasma volcanium} (\textit{A-Tvo}).

\newpage

\begin{longtable*}{@{\extracolsep{\fill}}lcccccccccccc}
\caption{\label{tab:virus} Information of \textit{virus}
{complete} sequences. } \\
\hline\hline
& \\
Virus &Length  & A$\%$ &C$\%$ &G$\%$ &T$\%$ &$\beta_A$ &$\beta_C$ &$\beta_G$ &$\beta_T$ &$S_{base}^{1}$ &$S_{beta}^{1}$ &Acc. No. \\
\hline
& \\
\endfirsthead

\caption{(Continued)}\\
\hline\hline
& \\
Virus &Length  & A$\%$ &C$\%$ &G$\%$ &T$\%$ &$\beta_A$ &$\beta_C$ &$\beta_G$ &$\beta_T$ &$S_{base}^{1}$ &$S_{beta}^{1}$ &Acc. No.\\
\hline
& \\
\endhead

&\\
\hline
\endfoot
&\\
\hline\hline
\endlastfoot

\textit{V-AsFV}     & 170101 & 0.304 & 0.194 & 0.195 & 0.307  & 0.926    & 0.890    & 0.909    & 0.964  &0.996  &0.985    & NC$\_$001659      \\
\textit{V-AsGV}     & 131680 & 0.309 & 0.182 & 0.191 & 0.318  & 0.947    & 0.938    & 1.058    & 1.069  &0.982  &0.940    & NC$\_$005839      \\
\textit{V-AmEPV}    & 232392 & 0.405 & 0.090 & 0.088 & 0.417  & 0.962    & 0.970    & 0.966    & 0.928  &0.986  &0.990    & NC$\_$002520      \\
\textit{V-AcNPV}    & 133894 & 0.293 & 0.203 & 0.204 & 0.300  & 0.984    & 0.952    & 1.019    & 0.949  &0.992  &0.974    & NC$\_$001623      \\
\textit{V-BoHV-1}   & 135301 & 0.135 & 0.359 & 0.365 & 0.140  & 0.899    & 1.093    & 1.043    & 0.925  &0.989  &0.981    & NC$\_$001847      \\
\textit{V-BoHV-5}   & 138390 & 0.124 & 0.372 & 0.376 & 0.128  & 1.025    & 1.047    & 0.964    & 1.028  &0.992  &0.979    & NC$\_$005261      \\
\textit{V-BPSV}     & 134431 & 0.178 & 0.322 & 0.323 & 0.177  & 1.029    & 0.933    & 0.895    & 1.118  &0.998  &0.968    & NC$\_$005337      \\
\textit{V-CalHV-3}  & 149696 & 0.262 & 0.247 & 0.245 & 0.245  & 0.805    & 1.044    & 1.025    & 0.838  &0.981  &0.986    & NC$\_$004367      \\
\textit{V-CMLV}     & 205719 & 0.336 & 0.166 & 0.166 & 0.332  & 0.941    & 1.009    & 0.987    & 0.962  &0.996  &0.989    & NC$\_$003391      \\
\textit{V-CNPV}     & 359853 & 0.352 & 0.152 & 0.152 & 0.344  & 0.967    & 0.984    & 1.044    & 0.918  &0.992  &0.972    & NC$\_$005309      \\
\textit{V-CeHV-1}   & 156789 & 0.128 & 0.369 & 0.375 & 0.127  & 1.074    & 1.048    & 1.043    & 1.055  &0.993  &0.994    & NC$\_$004812      \\
\textit{V-CCMV}     & 241087 & 0.192 & 0.307 & 0.310 & 0.191  & 0.996    & 0.886    & 0.869    & 0.971  &0.996  &0.989    & NC$\_$003521      \\
\textit{V-CfDEFNPV} & 131158 & 0.270 & 0.230 & 0.229 & 0.271  & 1.064    & 0.964    & 0.968    & 0.990  &0.998  &0.980    & NC$\_$005137      \\
\textit{V-CPXV}     & 224501 & 0.333 & 0.168 & 0.166 & 0.333  & 0.964    & 1.025    & 1.000    & 0.978  &0.998  &0.990    & NC$\_$003663      \\
\textit{V-EsV}      & 335593 & 0.244 & 0.258 & 0.260 & 0.238  & 0.991    & 0.999    & 1.170    & 0.950  &0.992  &0.948    & NC$\_$002687      \\
\textit{V-ECTV}     & 209771 & 0.335 & 0.167 & 0.165 & 0.334  & 0.952    & 1.022    & 0.977    & 0.969  &0.997  &0.984    & NC$\_$004105      \\
\textit{V-EHV-1}    & 150223 & 0.217 & 0.287 & 0.279 & 0.216  & 0.823    & 1.074    & 1.094    & 0.808  &0.991  &0.991    & NC$\_$001491      \\
\textit{V-EHV-2}    & 184427 & 0.216 & 0.293 & 0.282 & 0.209  & 0.941    & 0.907    & 0.875    & 0.975  &0.982  &0.982    & NC$\_$001650      \\
\textit{V-EHV-4}    & 145597 & 0.249 & 0.254 & 0.251 & 0.247  & 0.849    & 1.080    & 1.090    & 0.842  &0.995  &0.996    & NC$\_$001844      \\
\textit{V-FWPV}     & 288539 & 0.348 & 0.154 & 0.154 & 0.343  & 0.928    & 0.984    & 1.044    & 0.959  &0.995  &0.977    & NC$\_$002188      \\
\textit{V-GaHV-2}   & 138675 & 0.283 & 0.215 & 0.214 & 0.287  & 0.846    & 1.164    & 1.165    & 0.869  &0.995  &0.994    & NC$\_$002229      \\
\textit{V-GaHV-3}   & 164270 & 0.230 & 0.269 & 0.267 & 0.234  & 0.797    & 1.086    & 1.090    & 0.771  &0.994  &0.992    & NC$\_$002577      \\
\textit{V-GTPV}     & 149599 & 0.380 & 0.124 & 0.129 & 0.367  & 0.955    & 0.909    & 0.941    & 0.957  &0.982  &0.991    & NC$\_$004003      \\
\textit{V-HaNPV}    & 131403 & 0.301 & 0.194 & 0.196 & 0.309  & 0.946    & 0.961    & 0.978    & 0.988  &0.990  &0.985    & NC$\_$002654      \\
\textit{V-HzV-1}    & 228089 & 0.288 & 0.211 & 0.208 & 0.293  & 0.951    & 0.922    & 0.905    & 0.934  &0.992  &0.991    & NC$\_$004156      \\
\textit{V-HHV-1}    & 152261 & 0.159 & 0.338 & 0.345 & 0.158  & 0.833    & 1.203    & 1.181    & 0.820  &0.992  &0.991    & NC$\_$001806      \\
\textit{V-HHV-2}    & 154746 & 0.149 & 0.350 & 0.353 & 0.147  & 0.873    & 1.109    & 1.076    & 0.832  &0.995  &0.981    & NC$\_$001798      \\
\textit{V-HHV-4}    & 172281 & 0.198 & 0.305 & 0.295 & 0.203  & 0.877    & 0.949    & 1.117    & 0.958  &0.985  &0.936    & NC$\_$001345      \\
\textit{V-HHV-5}    & 230287 & 0.216 & 0.283 & 0.289 & 0.212  & 0.981    & 0.932    & 0.889    & 0.976  &0.990  &0.987    & NC$\_$001347      \\
\textit{V-HHV-6}    & 159321 & 0.289 & 0.217 & 0.208 & 0.287  & 0.876    & 0.946    & 0.924    & 0.868  &0.989  &0.992    & NC$\_$001664      \\
\textit{V-HHV-6B}   & 162114 & 0.287 & 0.217 & 0.211 & 0.286  & 0.859    & 0.955    & 0.965    & 0.828  &0.993  &0.989    & NC$\_$000898      \\
\textit{V-HHV-7}    & 144861 & 0.324 & 0.181 & 0.172 & 0.322  & 0.912    & 1.001    & 1.091    & 0.824  &0.989  &0.954    & NC$\_$001716      \\
\textit{V-HHV-8}    & 137508 & 0.237 & 0.275 & 0.260 & 0.228  & 1.091    & 1.021    & 1.046    & 1.011  &0.976  &0.975    & NC$\_$003409      \\
\textit{V-IcHV-1}   & 134226 & 0.214 & 0.281 & 0.281 & 0.224  & 0.887    & 0.940    & 1.001    & 1.075  &0.990  &0.936    & NC$\_$001493      \\
\textit{V-IIV-6}    & 212482 & 0.352 & 0.148 & 0.139 & 0.362  & 0.960    & 1.000    & 0.978    & 0.933  &0.981  &0.987    & NC$\_$003038      \\
\textit{V-LSDV}     & 150773 & 0.376 & 0.127 & 0.132 & 0.364  & 0.957    & 0.915    & 0.941    & 0.957  &0.983  &0.993    & NC$\_$003027      \\
\textit{V-LdNPV}    & 161046 & 0.213 & 0.287 & 0.288 & 0.213  & 0.969    & 0.919    & 0.954    & 0.964  &0.999  &0.989    & NC$\_$001973      \\
\textit{V-LCDV}     & 186250 & 0.363 & 0.135 & 0.138 & 0.365  & 0.951    & 1.023    & 1.136    & 0.986  &0.995  &0.964    & NC$\_$005902      \\
\textit{V-MMRV}     & 133719 & 0.245 & 0.267 & 0.258 & 0.230  & 0.896    & 0.982    & 1.177    & 0.971  &0.976  &0.933    & NC$\_$003401      \\
\textit{V-MacoNPV-A}& 155060 & 0.292 & 0.207 & 0.209 & 0.291  & 0.958    & 0.965    & 0.984    & 1.015  &0.997  &0.981    & NC$\_$003529      \\
\textit{V-MacoNPV-B}& 158482 & 0.302 & 0.199 & 0.202 & 0.298  & 0.978    & 0.960    & 0.985    & 0.988  &0.993  &0.991    & NC$\_$004117      \\
\textit{V-MsEPV}    & 236120 & 0.407 & 0.092 & 0.091 & 0.410  & 0.980    & 0.956    & 0.947    & 0.945  &0.996  &0.989    & NC$\_$001993      \\
\textit{V-MeHV-1}   & 159160 & 0.260 & 0.238 & 0.238 & 0.265  & 0.884    & 1.052    & 1.049    & 0.878  &0.995  &0.998    & NC$\_$002641      \\
\textit{V-MCV}      & 190289 & 0.184 & 0.315 & 0.318 & 0.182  & 1.021    & 0.929    & 0.930    & 0.985  &0.995  &0.990    & NC$\_$001731      \\
\textit{V-MPXV}     & 196858 & 0.335 & 0.166 & 0.165 & 0.334  & 0.942    & 1.001    & 0.996    & 0.956  &0.998  &0.995    & NC$\_$003310      \\
\textit{V-MCMV-1}   & 230278 & 0.204 & 0.292 & 0.295 & 0.209  & 1.024    & 0.939    & 0.883    & 0.981  &0.992  &0.974    & NC$\_$004065      \\
\textit{V-MYXV}     & 161773 & 0.287 & 0.217 & 0.219 & 0.278  & 0.941    & 1.026    & 1.018    & 0.950  &0.989  &0.996    & NC$\_$001132      \\
\textit{V-ORFV}     & 139962 & 0.184 & 0.318 & 0.316 & 0.181  & 1.021    & 0.877    & 0.812    & 1.069  &0.995  &0.970    & NC$\_$005336      \\
\textit{V-OpMNPV}   & 131995 & 0.223 & 0.276 & 0.275 & 0.225  & 0.958    & 0.907    & 0.976    & 1.238  &0.997  &0.914    & NC$\_$001875      \\
\textit{V-OsHV-1}   & 207439 & 0.314 & 0.192 & 0.195 & 0.298  & 0.889    & 1.033    & 1.017    & 0.909  &0.981  &0.991    & NC$\_$005881      \\
\textit{V-PBCV-1}   & 330743 & 0.300 & 0.201 & 0.198 & 0.300  & 0.967    & 1.063    & 1.062    & 0.931  &0.997  &0.991    & NC$\_$000852      \\
\textit{V-PsHV-1}   & 163025 & 0.193 & 0.308 & 0.301 & 0.198  & 0.898    & 1.060    & 1.070    & 0.937  &0.988  &0.988    & NC$\_$005264      \\
\textit{V-SFV}      & 159857 & 0.306 & 0.196 & 0.199 & 0.298  & 0.965    & 1.035    & 1.021    & 0.953  &0.989  &0.993    & NC$\_$001266      \\
\textit{V-RPXV}     & 197731 & 0.332 & 0.168 & 0.167 & 0.333  & 0.952    & 1.015    & 1.006    & 0.974  &0.998  &0.992    & NC$\_$005858      \\
\textit{V-RoMNPV}   & 131526 & 0.302 & 0.195 & 0.196 & 0.307  & 0.995    & 0.967    & 1.036    & 0.951  &0.994  &0.971    & NC$\_$004323      \\
\textit{V-RCMV}     & 230138 & 0.194 & 0.301 & 0.309 & 0.196  & 0.786    & 0.836    & 0.716    & 0.761  &0.990  &0.953    & NC$\_$002512      \\
\textit{V-SPPV}     & 149955 & 0.381 & 0.123 & 0.127 & 0.369  & 0.951    & 0.917    & 0.938    & 0.956  &0.984  &0.993    & NC$\_$004002      \\
\textit{V-SWSSV}    & 305107 & 0.302 & 0.205 & 0.205 & 0.288  & 0.980    & 0.972    & 1.170    & 0.918  &0.986  &0.936    & NC$\_$003225      \\
\textit{V-SpeiNPV}  & 135611 & 0.283 & 0.217 & 0.221 & 0.278  & 0.937    & 1.012    & 0.991    & 0.966  &0.991  &0.987    & NC$\_$002169      \\
\textit{V-SpltNPV}  & 139342 & 0.281 & 0.213 & 0.215 & 0.291  & 0.953    & 0.906    & 0.941    & 0.968  &0.988  &0.987    & NC$\_$003102      \\
\textit{V-SWPV}     & 146454 & 0.366 & 0.136 & 0.138 & 0.360  & 0.958    & 0.964    & 0.985    & 0.956  &0.992  &0.994    & NC$\_$003389      \\
\textit{V-TuHV-1}   & 195859 & 0.166 & 0.327 & 0.340 & 0.168  & 0.966    & 0.901    & 0.850    & 0.956  &0.985  &0.983    & NC$\_$002794      \\
\textit{V-VACV}     & 191737 & 0.333 & 0.167 & 0.167 & 0.333  & 0.986    & 0.963    & 0.984    & 0.984  &1.000  &0.994    & NC$\_$001559      \\
\textit{V-VARV}     & 185578 & 0.338 & 0.164 & 0.163 & 0.334  & 0.945    & 1.019    & 0.998    & 0.963  &0.995  &0.990    & NC$\_$001611      \\
\textit{V-XecnGV}   & 178733 & 0.297 & 0.202 & 0.205 & 0.296  & 0.983    & 0.974    & 0.935    & 0.954  &0.996  &0.982    & NC$\_$002331      \\
\textit{V-YMTV}     & 134721 & 0.355 & 0.148 & 0.150 & 0.347  & 0.941    & 1.177    & 1.216    & 0.949  &0.990  &0.989    & NC$\_$005179      \\
\textit{V-YLDV}     & 144575 & 0.367 & 0.134 & 0.136 & 0.363  & 0.952    & 1.153    & 1.150    & 0.946  &0.994  &0.998    & NC$\_$002642      \\
\end{longtable*}

\begin{longtable*}{@{\extracolsep{\fill}}llcccccccccccr}
\caption{\label{tab:bacteria}Information of Bacteria. The Bergey
Code is a shorthand of the lineage of the organism according to
the order: phylum, class, order, family, genus. For
species/strains (sp/str) belong to the fourteenth phylum, the
subclass and suborder is also given. Items are ordered by the
Bergey code, so species/strains closely related are listed together. } \\

\hline\hline
& \\
Beygey code & Sp/str &Length & A$\%$ &C$\%$ &G$\%$ &T$\%$ &$\beta_A$ &$\beta_C$ &$\beta_G$ & $\beta_T$ & $S_{base}^{1}$ & $S_{beta}^{1}$ &Acc. No.\\
\hline
& \\
\endfirsthead

\caption{(Continued)}\\
\hline\hline
& \\
Beygey code & Sp/str &Length & A$\%$ &C$\%$ &G$\%$ &T$\%$ &$\beta_A$ &$\beta_C$ &$\beta_G$ & $\beta_T$ & $S_{base}^{1}$  & $S_{beta}^{1}$  &Acc. No. \\
\hline
& \\
\endhead

&\\
\hline
\endfoot
&\\
\hline\hline %
\endlastfoot

B.1.1.1.1.1           &  \textit{B-Aae}  & 1551335 & 0.284  & 0.217  & 0.218  & 0.281 & 0.888 & 0.976 & 0.976 & 0.901 & 0.996 & 0.997   & NC$\_$000918\\
B.2.1.1.1.1           &  \textit{B-Tma}  & 1860725 & 0.270  & 0.228  & 0.235  & 0.268 & 0.907 & 0.951 & 0.928 & 0.905 & 0.991 & 0.993   & NC$\_$000853\\
B.4.1.1.1.1           &  \textit{B-Dra1} & 2648638 & 0.165  & 0.335  & 0.335  & 0.165 & 1.118 & 0.982 & 0.962 & 1.053 & 1.000 & 0.979   & NC$\_$001263\\
                      &  \textit{B-Dra2} &  412348 & 0.170  & 0.333  & 0.334  & 0.164 & 1.132 & 0.877 & 0.925 & 1.173 & 0.993 & 0.978   & NC$\_$001264\\
B.10.l                &  \textit{B-Tel}  & 2593857 & 0.231  & 0.269  & 0.270  & 0.230 & 1.042 & 0.970 & 0.993 & 1.012 & 0.998 & 0.987   & NC$\_$004113\\
B.10.1.1.1.11         &  \textit{B-Pma1} & 1751080 & 0.319  & 0.182  & 0.182  & 0.317 & 0.971 & 0.938 & 1.037 & 0.978 & 0.998 & 0.973   & NC$\_$005042\\
                      &  \textit{B-Pma2} & 1657990 & 0.345  & 0.155  & 0.153  & 0.347 & 0.958 & 0.928 & 1.062 & 0.957 & 0.996 & 0.965   & NC$\_$005072\\
                      &  \textit{B-Pma3} & 2410873 & 0.256  & 0.262  & 0.245  & 0.236 & 0.974 & 0.884 & 0.934 & 0.977 & 0.963 & 0.986   & NC$\_$005071\\
B.10.1.1.1.13         &  \textit{B-SspW} & 2434428 & 0.202  & 0.297  & 0.297  & 0.204 & 1.077 & 0.843 & 0.822 & 1.085 & 0.998 & 0.992   & NC$\_$005070\\
B.10.1.1.1.14         &  \textit{B-SspP} & 3573470 & 0.261  & 0.238  & 0.239  & 0.262 & 1.055 & 0.938 & 0.927 & 1.068 & 0.998 & 0.994   & NC$\_$000911\\
B.10.1.4.1.8          &  \textit{B-Nsp}  & 6413771 & 0.293  & 0.206  & 0.207  & 0.294 & 0.985 & 0.989 & 1.006 & 0.975 & 0.998 & 0.993   & NC$\_$003272\\
B.11.1.1.1.1          &  \textit{B-Cte}  & 2154946 & 0.219  & 0.284  & 0.281  & 0.216 & 1.062 & 0.929 & 0.923 & 1.057 & 0.994 & 0.997   & NC$\_$002932\\
B.12.1.2.1.1          &  \textit{B-Rco}  & 1268755 & 0.337  & 0.161  & 0.163  & 0.339 & 0.978 & 0.995 & 1.007 & 0.971 & 0.996 & 0.995   & NC$\_$003103\\
                      &  \textit{B-Rpr}  & 1111523 & 0.354  & 0.144  & 0.146  & 0.356 & 0.955 & 1.023 & 1.018 & 0.988 & 0.996 & 0.990   & NC$\_$000963\\
B.12.1.5.1.1          &  \textit{B-Ccr}  & 4016947 & 0.165  & 0.337  & 0.335  & 0.163 & 0.986 & 0.994 & 0.976 & 1.030 & 0.996 & 0.984   & NC$\_$002696\\
B.12.1.6.1.2          &  \textit{B-Atu1} & 2841490 & 0.205  & 0.300  & 0.294  & 0.202 & 1.018 & 0.978 & 0.977 & 0.979 & 0.991 & 0.990   & NC$\_$003304\\
                      &  \textit{B-Atu2} & 2075560 & 0.203  & 0.297  & 0.296  & 0.204 & 1.038 & 0.959 & 0.965 & 1.017 & 0.998 & 0.993   & NC$\_$003305\\
B.12.1.6.1.6          &  \textit{B-Sme1} & 3654135 & 0.186  & 0.315  & 0.312  & 0.186 & 0.975 & 0.971 & 0.971 & 1.002 & 0.997 & 0.993   & NC$\_$003047\\
                      &  \textit{B-Sme2} & 1354226 & 0.200  & 0.303  & 0.301  & 0.197 & 0.971 & 0.992 & 0.988 & 0.973 & 0.995 & 0.998   & NC$\_$003037\\
                      &  \textit{B-Sme3} & 1683333 & 0.188  & 0.311  & 0.313  & 0.188 & 0.991 & 0.975 & 0.991 & 0.974 & 0.998 & 0.992   & NC$\_$003078\\
B.12.1.6.3.1          &  \textit{B-Bme1} & 2117144 & 0.214  & 0.285  & 0.287  & 0.215 & 1.011 & 0.963 & 0.943 & 1.005 & 0.997 & 0.993   & NC$\_$003317\\
                      &  \textit{B-Bme2} & 1177787 & 0.214  & 0.286  & 0.288  & 0.213 & 1.006 & 0.968 & 0.942 & 0.970 & 0.997 & 0.984   & NC$\_$003318\\
                      &  \textit{B-Bsu1} & 2107792 & 0.214  & 0.287  & 0.285  & 0.214 & 1.005 & 0.946 & 0.965 & 1.006 & 0.998 & 0.995   & NC$\_$004310\\
                      &  \textit{B-Bsu2} & 1207381 & 0.213  & 0.287  & 0.286  & 0.214 & 0.970 & 0.943 & 0.970 & 1.006 & 0.998 & 0.984   & NC$\_$004311\\
B.12.1.6.4.6          &  \textit{B-Mlo}  & 7036074 & 0.186  & 0.316  & 0.311  & 0.186 & 1.000 & 0.973 & 0.970 & 0.992 & 0.995 & 0.997   & NC$\_$002678\\
B.12.1.6.7.1          &  \textit{B-Bja}  & 9105828 & 0.180  & 0.320  & 0.320  & 0.180 & 0.967 & 0.986 & 0.980 & 0.969 & 1.000 & 0.998   & NC$\_$004463\\
B.12.2.1.2.1          &  \textit{B-Rso}  & 3716413 & 0.164  & 0.333  & 0.337  & 0.166 & 1.032 & 0.946 & 0.948 & 1.062 & 0.994 & 0.992   & NC$\_$003295\\
B.12.2.1.4.3          &  \textit{B-Bbr}  & 5339179 & 0.159  & 0.339  & 0.342  & 0.160 & 1.000 & 0.950 & 0.978 & 1.060 & 0.996 & 0.978   & NC$\_$002927\\
                      &  \textit{B-Bpa}  & 4773551 & 0.159  & 0.338  & 0.343  & 0.160 & 1.020 & 0.952 & 0.982 & 1.061 & 0.994 & 0.982   & NC$\_$002928\\
                      &  \textit{B-Bpe}  & 4086189 & 0.161  & 0.337  & 0.340  & 0.161 & 1.011 & 0.972 & 0.963 & 1.013 & 0.997 & 0.997   & NC$\_$002929\\
B.12.2.4.1.1          &  \textit{B-Nme1} & 2272351 & 0.242  & 0.256  & 0.260  & 0.243 & 0.969 & 0.917 & 0.950 & 0.944 & 0.995 & 0.985   & NC$\_$003112\\
                      &  \textit{B-Nme2} & 2184406 & 0.240  & 0.259  & 0.259  & 0.242 & 0.970 & 0.934 & 0.944 & 0.962 & 0.998 & 0.995   & NC$\_$003116\\
B.12.2.4.1.5          &  \textit{B-Cvi}  & 4751080 & 0.175  & 0.324  & 0.324  & 0.176 & 1.109 & 0.881 & 0.884 & 1.125 & 0.999 & 0.995   & NC$\_$005085\\
B.12.3.3.1.1          &  \textit{B-Xax}  & 5175554 & 0.176  & 0.324  & 0.323  & 0.176 & 1.037 & 0.958 & 0.951 & 1.071 & 0.999 & 0.990   & NC$\_$003919\\
                      &  \textit{B-Xca}  & 5076188 & 0.175  & 0.325  & 0.325  & 0.174 & 1.056 & 0.941 & 0.958 & 1.049 & 0.999 & 0.994   & NC$\_$003902\\
B.12.3.3.1.9          &  \textit{B-Xfa}  & 2679306 & 0.225  & 0.249  & 0.277  & 0.248 & 0.962 & 1.062 & 1.051 & 0.946 & 0.949 & 0.993   & NC$\_$002488\\
B.12.3.6.2.1          &  \textit{B-Cbu}  & 1995275 & 0.287  & 0.213  & 0.213  & 0.286 & 1.003 & 0.967 & 0.994 & 0.982 & 0.999 & 0.988   & NC$\_$002971\\
B.12.3.9.1.1          &  \textit{B-Pae}  & 6264403 & 0.169  & 0.336  & 0.330  & 0.166 & 1.107 & 0.932 & 0.928 & 1.100 & 0.991 & 0.997   & NC$\_$002516\\
                      &  \textit{B-Ppu}  & 6181863 & 0.192  & 0.306  & 0.310  & 0.193 & 1.103 & 0.929 & 0.925 & 1.034 & 0.995 & 0.982   & NC$\_$002947\\
                      &  \textit{B-Psy}  & 6397126 & 0.208  & 0.292  & 0.292  & 0.208 & 1.027 & 0.935 & 0.928 & 1.010 & 1.000 & 0.994   & NC$\_$004578\\
B.12.3.10.1.7         &  \textit{B-Son}  & 4969803 & 0.270  & 0.230  & 0.230  & 0.270 & 1.009 & 0.999 & 0.969 & 0.992 & 1.000 & 0.988   & NC$\_$004347\\
B.12.3.11.1.1         &  \textit{B-Vch1} & 2961149 & 0.260  & 0.238  & 0.239  & 0.263 & 1.040 & 0.919 & 0.961 & 1.033 & 0.996 & 0.988   & NC$\_$002505\\
                      &  \textit{B-Vch2} & 1072315 & 0.265  & 0.233  & 0.236  & 0.266 & 1.008 & 0.920 & 0.956 & 0.927 & 0.996 & 0.969   & NC$\_$002506\\
                      &  \textit{B-Vvu1} & 3281945 & 0.267  & 0.231  & 0.233  & 0.269 & 1.042 & 0.949 & 0.946 & 0.970 & 0.996 & 0.981   & NC$\_$004459\\
                      &  \textit{B-Vvu2} & 1844853 & 0.264  & 0.236  & 0.235  & 0.265 & 0.999 & 0.935 & 0.938 & 1.001 & 0.998 & 0.999   & NC$\_$004460\\
                      &  \textit{B-Vpa1} & 3288558 & 0.272  & 0.227  & 0.227  & 0.274 & 0.972 & 0.988 & 0.974 & 0.988 & 0.998 & 0.992   & NC$\_$004603\\
                      &  \textit{B-Vpa2} & 1877212 & 0.272  & 0.227  & 0.226  & 0.274 & 0.983 & 0.982 & 0.938 & 0.979 & 0.997 & 0.988   & NC$\_$004605\\
B.12.3.13.1.5         &  \textit{B-BapS} &  641454 & 0.375  & 0.125  & 0.128  & 0.372 & 0.942 & 1.082 & 1.158 & 0.984 & 0.994 & 0.972   & NC$\_$004061\\
                      &  \textit{B-Bsp}  &  640681 & 0.371  & 0.131  & 0.132  & 0.366 & 0.954 & 1.085 & 1.151 & 0.978 & 0.994 & 0.978   & NC$\_$002528\\
                      &  \textit{B-Bap}  &  615980 & 0.371  & 0.127  & 0.127  & 0.375 & 0.922 & 1.016 & 1.056 & 0.970 & 0.996 & 0.978   & NC$\_$004545\\
B.12.3.13.1.13        &  \textit{B-Eco1} & 5231428 & 0.248  & 0.253  & 0.252  & 0.247 & 1.021 & 0.918 & 0.923 & 1.009 & 0.998 & 0.996   & NC$\_$004431\\
                      &  \textit{B-Eco2} & 4639221 & 0.246  & 0.254  & 0.254  & 0.246 & 1.023 & 0.925 & 0.934 & 1.005 & 1.000 & 0.993   & NC$\_$000913\\
                      &  \textit{B-Eco3} & 5498450 & 0.248  & 0.252  & 0.253  & 0.247 & 1.028 & 0.918 & 0.934 & 1.025 & 0.998 & 0.995   & NC$\_$002695\\
                      &  \textit{B-Eco4} & 5528445 & 0.248  & 0.252  & 0.252  & 0.247 & 1.028 & 0.920 & 0.927 & 1.026 & 0.999 & 0.998   & NC$\_$002655\\
B.12.3.13.1.32        &  \textit{B-Sty1} & 4809037 & 0.239  & 0.260  & 0.261  & 0.240 & 1.014 & 0.922 & 0.929 & 1.010 & 0.998 & 0.997   & NC$\_$003198\\
                      &  \textit{B-Sty2} & 4857432 & 0.239  & 0.261  & 0.261  & 0.239 & 1.024 & 0.928 & 0.925 & 1.003 & 1.000 & 0.994   & NC$\_$003197\\
                      &  \textit{B-Sty3} & 4791961 & 0.239  & 0.260  & 0.261  & 0.240 & 1.025 & 0.912 & 0.930 & 1.015 & 0.998 & 0.993   & NC$\_$004631\\
B.12.3.13.1.34        &  \textit{B-Sfl}  & 4607203 & 0.246  & 0.255  & 0.254  & 0.245 & 1.043 & 0.951 & 0.931 & 1.020 & 0.998 & 0.989   & NC$\_$004337\\
B.12.3.13.1.38        &  \textit{B-Wbr}  &  697724 & 0.388  & 0.113  & 0.112  & 0.387 & 0.906 & 1.044 & 1.083 & 0.920 & 0.998 & 0.987   & NC$\_$004344\\
B.12.3.13.1.40        &  \textit{B-Ype1} & 4653728 & 0.262  & 0.237  & 0.239  & 0.262 & 1.024 & 0.937 & 0.944 & 1.018 & 0.998 & 0.997   & NC$\_$003143\\
                      &  \textit{B-Ype2} & 4600755 & 0.261  & 0.237  & 0.239  & 0.263 & 1.023 & 0.952 & 0.930 & 1.021 & 0.996 & 0.994   & NC$\_$004088\\
B.12.3.14.1.1         &  \textit{B-Pmu}  & 2257487 & 0.299  & 0.199  & 0.205  & 0.297 & 1.005 & 0.979 & 0.997 & 1.001 & 0.992 & 0.994   & NC$\_$002663\\
B.12.3.14.1.3         &  \textit{B-Hin}  & 1830138 & 0.310  & 0.192  & 0.190  & 0.308 & 0.979 & 1.045 & 1.051 & 0.990 & 0.996 & 0.996   & NC$\_$000907\\
                      &  \textit{B-Hdu}  & 1698955 & 0.305  & 0.185  & 0.197  & 0.312 & 0.980 & 1.015 & 1.025 & 0.949 & 0.981 & 0.990   & NC$\_$002940\\
B.12.5.1.1.1          &  \textit{B-Cje}  & 1641481 & 0.348  & 0.153  & 0.152  & 0.346 & 0.981 & 0.880 & 0.960 & 0.965 & 0.997 & 0.975   & NC$\_$002163\\
B.12.5.1.2.1          &  \textit{B-Hpy1} & 1667867 & 0.303  & 0.196  & 0.193  & 0.308 & 0.981 & 0.956 & 0.950 & 1.016 & 0.992 & 0.989   & NC$\_$000915\\
                      &  \textit{B-Hpy2} & 1643831 & 0.303  & 0.197  & 0.195  & 0.305 & 0.983 & 0.957 & 0.948 & 0.996 & 0.996 & 0.994   & NC$\_$000921\\
                      &  \textit{B-Hhe}  & 1799146 & 0.322  & 0.182  & 0.177  & 0.319 & 0.996 & 0.946 & 0.931 & 0.997 & 0.992 & 0.996   & NC$\_$004917\\
B.12.5.1.2.2          &  \textit{B-Wsu}  & 2110355 & 0.257  & 0.239  & 0.245  & 0.259 & 1.083 & 0.914 & 0.947 & 1.026 & 0.992 & 0.977   & NC$\_$005090\\
B.13.1.1.1.1          &  \textit{B-Cac}  & 3940880 & 0.346  & 0.154  & 0.155  & 0.345 & 0.891 & 0.807 & 0.965 & 0.896 & 0.998 & 0.954   & NC$\_$003030\\
                      &  \textit{B-Cpe}  & 3031430 & 0.350  & 0.147  & 0.138  & 0.365 & 0.867 & 0.913 & 1.015 & 0.858 & 0.976 & 0.970   & NC$\_$003366\\
                      &  \textit{B-Cte}  & 2799251 & 0.353  & 0.146  & 0.141  & 0.359 & 0.870 & 0.952 & 0.925 & 0.830 & 0.989 & 0.981   & NC$\_$004557\\
B.13.1.2.1.8          &  \textit{B-Tte}  & 2689445 & 0.312  & 0.188  & 0.188  & 0.313 & 0.938 & 0.876 & 0.912 & 0.930 & 0.999 & 0.988   & NC$\_$003869\\
B.13.2.1.1.1          &  \textit{B-Mge}  &  580074 & 0.346  & 0.158  & 0.159  & 0.337 & 0.978 & 0.984 & 1.072 & 0.930 & 0.990 & 0.966   & NC$\_$000908\\
                      &  \textit{B-Mpe}  & 1358633 & 0.370  & 0.128  & 0.129  & 0.372 & 0.945 & 1.179 & 0.948 & 0.951 & 0.997 & 0.941   & NC$\_$004432\\
                      &  \textit{B-Mpn}  &  816394 & 0.305  & 0.200  & 0.201  & 0.295 & 0.990 & 1.017 & 0.998 & 0.970 & 0.989 & 0.990   & NC$\_$000912\\
                      &  \textit{B-Mpu}  &  963879 & 0.370  & 0.133  & 0.133  & 0.364 & 0.917 & 1.195 & 1.261 & 0.918 & 0.994 & 0.984   & NC$\_$002771\\
                      &  \textit{B-Mga}  &  996422 & 0.345  & 0.157  & 0.157  & 0.341 & 0.959 & 1.010 & 1.158 & 0.950 & 0.996 & 0.961   & NC$\_$004829\\
B.13.2.1.1.4          &  \textit{B-Uur}  &  751719 & 0.373  & 0.126  & 0.129  & 0.372 & 0.954 & 0.993 & 1.130 & 0.944 & 0.996 & 0.963   & NC$\_$002162\\
B.13.3.1.1            &  \textit{B-Oih}  & 3630528 & 0.321  & 0.179  & 0.178  & 0.322 & 0.943 & 0.924 & 1.009 & 0.934 & 0.998 & 0.975   & NC$\_$004193\\
B.13.3.1.1.1          &  \textit{B-Ban}  & 5093554 & 0.323  & 0.178  & 0.175  & 0.325 & 0.939 & 0.865 & 0.890 & 0.951 & 0.995 & 0.990   & NC$\_$003995\\
                      &  \textit{B-Bha}  & 4202353 & 0.282  & 0.217  & 0.220  & 0.281 & 0.966 & 0.912 & 0.932 & 0.971 & 0.996 & 0.993   & NC$\_$002570\\
                      &  \textit{B-Bsu}  & 4214814 & 0.282  & 0.218  & 0.217  & 0.283 & 0.941 & 0.931 & 0.957 & 0.941 & 0.998 & 0.993   & NC$\_$000964\\
B.13.3.1.4.1          &  \textit{B-Lin}  & 3011208 & 0.313  & 0.189  & 0.186  & 0.313 & 0.930 & 1.061 & 1.006 & 0.924 & 0.997 & 0.984   & NC$\_$003212\\
                      &  \textit{B-Lmo}  & 2944528 & 0.310  & 0.191  & 0.189  & 0.310 & 0.942 & 1.057 & 1.000 & 0.933 & 0.998 & 0.983   & NC$\_$003210\\
B.13.3.1.5.1          &  \textit{B-Sau1} & 2878040 & 0.335  & 0.164  & 0.165  & 0.337 & 0.927 & 1.069 & 0.999 & 0.956 & 0.997 & 0.975   & NC$\_$002758\\
                      &  \textit{B-Sau2} & 2820462 & 0.334  & 0.164  & 0.164  & 0.338 & 0.924 & 1.068 & 1.028 & 0.959 & 0.996 & 0.981   & NC$\_$003923\\
                      &  \textit{B-Sau3} & 2814816 & 0.334  & 0.164  & 0.164  & 0.337 & 0.929 & 1.073 & 1.003 & 0.959 & 0.997 & 0.975   & NC$\_$002745\\
                      &  \textit{B-Sep}  & 2499279 & 0.335  & 0.162  & 0.159  & 0.344 & 0.945 & 1.148 & 0.934 & 0.966 & 0.988 & 0.941   & NC$\_$004461\\
B.13.3.2.1.1          &  \textit{B-Lpl}  & 3308274 & 0.277  & 0.223  & 0.222  & 0.278 & 0.986 & 0.882 & 0.907 & 0.980 & 0.998 & 0.992   & NC$\_$004567\\
B.13.3.2.6.1          &  \textit{B-Sag1} & 2160267 & 0.323  & 0.179  & 0.178  & 0.321 & 0.980 & 0.920 & 1.095 & 0.962 & 0.997 & 0.951   & NC$\_$004116\\
                      &  \textit{B-Sag2} & 2211485 & 0.323  & 0.178  & 0.178  & 0.321 & 0.982 & 0.918 & 1.103 & 0.969 & 0.998 & 0.950   & NC$\_$004368\\
                      &  \textit{B-Smu}  & 2030921 & 0.315  & 0.185  & 0.183  & 0.317 & 0.986 & 1.068 & 1.028 & 0.988 & 0.996 & 0.990   & NC$\_$004350\\
                      &  \textit{B-Spn1} & 2038615 & 0.302  & 0.198  & 0.199  & 0.301 & 0.993 & 0.905 & 0.890 & 0.999 & 0.998 & 0.994   & NC$\_$003098\\
                      &  \textit{B-Spn2} & 2160837 & 0.303  & 0.198  & 0.199  & 0.300 & 0.995 & 0.890 & 1.016 & 0.995 & 0.996 & 0.968   & NC$\_$003028\\
                      &  \textit{B-Spy1} & 1895017 & 0.307  & 0.192  & 0.193  & 0.308 & 0.970 & 0.970 & 1.036 & 0.979 & 0.998 & 0.981   & NC$\_$003485\\
                      &  \textit{B-Spy2} & 1900521 & 0.305  & 0.194  & 0.192  & 0.309 & 0.972 & 0.971 & 1.067 & 0.975 & 0.994 & 0.975   & NC$\_$004070\\
                      &  \textit{B-Spy3} & 1852441 & 0.309  & 0.191  & 0.194  & 0.306 & 0.977 & 0.973 & 1.034 & 0.976 & 0.994 & 0.984   & NC$\_$002737\\
                      &  \textit{B-Spy4} & 1894275 & 0.309  & 0.190  & 0.195  & 0.306 & 0.978 & 0.961 & 1.014 & 0.972 & 0.992 & 0.985   & NC$\_$004606\\
B.13.3.2.6.2          &  \textit{B-Lla}  & 2365589 & 0.324  & 0.176  & 0.178  & 0.323 & 0.956 & 1.090 & 0.973 & 0.985 & 0.997 & 0.964   & NC$\_$002662\\
B.14.(1.5).(1.7).1.1  &  \textit{B-Cef}  & 3147090 & 0.184  & 0.315  & 0.316  & 0.185 & 1.038 & 0.966 & 0.959 & 1.040 & 0.998 & 0.998   & NC$\_$004369\\
                      &  \textit{B-Cgl}  & 3309401 & 0.231  & 0.270  & 0.268  & 0.231 & 0.953 & 0.980 & 1.003 & 0.960 & 0.998 & 0.992   & NC$\_$003450\\
B.14.(1.5).(1.7).4.1  &  \textit{B-Mle}  & 3268203 & 0.210  & 0.287  & 0.291  & 0.212 & 0.971 & 0.959 & 0.957 & 1.005 & 0.994 & 0.991   & NC$\_$002677\\
                      &  \textit{B-Mtu1} & 4403836 & 0.172  & 0.329  & 0.327  & 0.172 & 0.962 & 1.134 & 1.159 & 0.975 & 0.998 & 0.991   & NC$\_$002755\\
                      &  \textit{B-Mtu2} & 4411529 & 0.172  & 0.329  & 0.327  & 0.172 & 0.964 & 1.133 & 1.163 & 0.974 & 0.998 & 0.991   & NC$\_$000962\\
                      &  \textit{B-Mbo}  & 4345492 & 0.172  & 0.329  & 0.327  & 0.172 & 0.973 & 1.135 & 1.169 & 0.976 & 0.998 & 0.991   & NC$\_$002945\\
B.14.(1.5).(1.11).1.1 &  \textit{B-Sco}  & 8667507 & 0.139  & 0.360  & 0.361  & 0.140 & 1.007 & 0.987 & 0.985 & 1.011 & 0.998 & 0.998   & NC$\_$003888\\
                      &  \textit{B-Sav}  & 9025608 & 0.147  & 0.354  & 0.353  & 0.146 & 1.020 & 1.034 & 0.985 & 1.016 & 0.998 & 0.987   & NC$\_$003155\\
B.14.(1.5).2.1.1      &  \textit{B-Blo}  & 2256646 & 0.200  & 0.301  & 0.301  & 0.199 & 1.126 & 0.974 & 0.947 & 1.045 & 0.999 & 0.974   & NC$\_$004307\\
B.15.1.1.1.4          &  \textit{B-Psp}  & 7145576 & 0.224  & 0.279  & 0.275  & 0.222 & 0.990 & 0.987 & 1.016 & 0.989 & 0.994 & 0.992   & NC$\_$005027\\
B.16.1.1.1.1          &  \textit{B-Cmu}  & 1072950 & 0.299  & 0.201  & 0.202  & 0.298 & 0.976 & 0.899 & 0.957 & 0.990 & 0.998 & 0.981   & NC$\_$002620\\
                      &  \textit{B-Ctr}  & 1042519 & 0.294  & 0.206  & 0.207  & 0.293 & 0.976 & 0.893 & 0.959 & 0.997 & 0.998 & 0.977   & NC$\_$000117\\
B.16.1.1.1.2          &  \textit{B-Cpn1} & 1229858 & 0.296  & 0.203  & 0.203  & 0.299 & 0.996 & 1.000 & 0.955 & 0.983 & 0.997 & 0.985   & NC$\_$002179\\
                      &  \textit{B-Cpn2} & 1230230 & 0.299  & 0.203  & 0.203  & 0.296 & 0.983 & 0.954 & 1.000 & 0.996 & 0.997 & 0.985   & NC$\_$000922\\
                      &  \textit{B-Cpn3} & 1226565 & 0.299  & 0.203  & 0.203  & 0.296 & 0.984 & 0.954 & 0.998 & 0.995 & 0.997 & 0.986   & NC$\_$002491\\
                      &  \textit{B-Cpn4} & 1225935 & 0.304  & 0.196  & 0.196  & 0.303 & 0.984 & 0.954 & 0.999 & 0.995 & 0.996 & 0.986   & NC$\_$005043\\
                      &  \textit{B-Cca}  & 1173390 & 0.299  & 0.203  & 0.202  & 0.296 & 0.988 & 0.977 & 0.985 & 0.991 & 0.999 & 0.997   & NC$\_$003361\\
B.17.1.1.1.2          &  \textit{B-Bbu}  &  910724 & 0.355  & 0.144  & 0.142  & 0.359 & 0.952 & 0.964 & 0.869 & 0.966 & 0.994 & 0.971   & NC$\_$001318\\
B.17.1.1.1.9          &  \textit{B-Tpa}  & 1138011 & 0.235  & 0.262  & 0.266  & 0.237 & 0.974 & 0.903 & 0.898 & 0.959 & 0.994 & 0.995   & NC$\_$000919\\
B.17.1.1.3.2          &  \textit{B-Lin1} & 4332241 & 0.325  & 0.174  & 0.176  & 0.325 & 0.963 & 0.977 & 0.972 & 0.959 & 0.998 & 0.998   & NC$\_$004342\\
                      &  \textit{B-Lin2} &  358943 & 0.324  & 0.175  & 0.177  & 0.325 & 0.998 & 0.981 & 0.960 & 0.938 & 0.997 & 0.979   & NC$\_$004343\\
B.20.1.1.1.1          &  \textit{B-Bth}  & 6260361 & 0.285  & 0.213  & 0.215  & 0.287 & 0.957 & 0.896 & 0.900 & 0.945 & 0.996 & 0.996   & NC$\_$004663\\
B.20.1.1.3.1          &  \textit{B-Pgi}  & 2343476 & 0.258  & 0.241  & 0.242  & 0.259 & 1.010 & 0.925 & 0.934 & 1.012 & 0.998 & 0.997   & NC$\_$002950\\
B.21.1.1.1.1          &  \textit{B-Fnu}  & 2174500 & 0.358  & 0.140  & 0.132  & 0.370 & 0.887 & 0.873 & 0.846 & 0.895 & 0.980 & 0.990   & NC$\_$003454\\
\end{longtable*}

\begin{table*}
\caption{\label{tab:archaea}Information of Archaea. The Bergey
code is the lineage of the organism according to the order:
phylum, class, order, family, genus. Items are ordered by the
Bergey code, so species/strains closely related are listed
together.}
\begin{ruledtabular}
\begin{tabular}{llcccccccccccr}
Beygey code  & Sp/str   & Length    & A$\%$  & C$\%$  & G$\%$  & T$\%$   &$\beta_A$ &$\beta_C$ &$\beta_G$ &$\beta_T$ &$S_{base}^{1}$ &$S_{beta}^{1}$ & Acc. No.\\
\colrule
 A.1.1.2.1.3 &\textit{A-Ape}   & 1669695   & 0.216  & 0.284  & 0.280  & 0.221   & 0.999     &  0.933    &  0.914    &  1.024 & 0.991          & 0.989        & NC$\_$000854 \\
 A.1.1.3.1.1 &\textit{A-Sso}   & 2992245   & 0.319  & 0.179  & 0.179  & 0.323   & 0.947     &  0.923    &  0.955    &  0.944 & 0.996          & 0.991        & NC$\_$002754 \\
             &\textit{A-Sto}   & 2694756   & 0.334  & 0.163  & 0.165  & 0.338   & 0.965     &  1.037    &  0.944    &  0.950 & 0.994          & 0.972        & NC$\_$003106 \\
 A.2.1.1.1.1 &\textit{A-Mth}   & 1751377   & 0.251  & 0.247  & 0.248  & 0.254   & 0.958     &  0.936    &  0.939    &  0.986 & 0.996          & 0.992        & NC$\_$000916 \\
 A.2.6.1.1.1 &\textit{A-Afu}   & 2178400   & 0.258  & 0.242  & 0.244  & 0.256   & 0.967     &  0.910    &  0.885    &  0.992 & 0.996          & 0.987        & NC$\_$000917 \\
 A.2.3.1.1.1 &\textit{A-Hsp}   & 2014239   & 0.161  & 0.340  & 0.339  & 0.160   & 1.007     &  0.950    &  0.949    &  1.030 & 0.998          & 0.994        & NC$\_$002607 \\
 A.2.2.1.1.1 &\textit{A-Mja}   & 1664970   & 0.344  & 0.155  & 0.159  & 0.341   & 0.919     &  0.916    &  0.909    &  0.927 & 0.993          & 0.996        & NC$\_$000909 \\
 A.2.7.1.1.1 &\textit{A-Mka}   & 1694969   & 0.195  & 0.307  & 0.304  & 0.194   & 1.106     &  0.943    &  0.951    &  1.065 & 0.996          & 0.988        & NC$\_$003551 \\
 A.2.2.3.1.1 &\textit{A-Mac}   & 5751492   & 0.285  & 0.214  & 0.213  & 0.288   & 0.937     &  0.917    &  0.911    &  0.940 & 0.996          & 0.998        & NC$\_$003552 \\
 A.2.2.3.1.1 &\textit{A-Mma}   & 4096345   & 0.293  & 0.207  & 0.208  & 0.292   & 0.939     &  0.925    &  0.930    &  0.944 & 0.998          & 0.997        & NC$\_$003901 \\
 A.2.5.1.1.3 &\textit{A-Pab}   & 1765118   & 0.276  & 0.224  & 0.223  & 0.277   & 0.925     &  0.886    &  0.911    &  0.926 & 0.998          & 0.993        & NC$\_$000868 \\
             &\textit{A-Pfu}   & 1908256   & 0.296  & 0.204  & 0.204  & 0.296   & 0.924     &  0.897    &  0.953    &  0.914 & 1.000          & 0.982        & NC$\_$003413 \\
             &\textit{A-Pho}   & 1738505   & 0.290  & 0.212  & 0.207  & 0.291   & 0.938     &  0.967    &  0.922    &  0.933 & 0.994          & 0.984        & NC$\_$000961 \\
 A.2.4.1.1.1 &\textit{A-Tac}   & 1564906   & 0.272  & 0.229  & 0.231  & 0.268   & 0.949     &  0.947    &  0.954    &  0.944 & 0.994          & 0.997        & NC$\_$002578 \\
             &\textit{A-Tvo}   & 1584804   & 0.302  & 0.199  & 0.200  & 0.299   & 0.941     &  0.946    &  0.950    &  0.965 & 0.996          & 0.993        & NC$\_$002689 \\
\end{tabular}
\end{ruledtabular}
\end{table*}

\begin{table*}
\caption{\label{tab:scer}Information of the \textit{Saccharomyces
cerevisiae} genome.}
\begin{ruledtabular}
\begin{tabular}{lccccccccccc}
Chromosome &Length   & A$\%$  & C$\%$  & G$\%$  & T$\%$  &$\beta_A$ &$\beta_C$ &$\beta_G$ &$\beta_T$ &$S_{base}^{1}$ &$S_{beta}^{1}$\\
\colrule
 chr1      & 230203  & 0.303  & 0.194  & 0.199  & 0.304  & 0.957    & 1.072    & 1.034    & 0.924  & 0.994      & 0.982  \\
 chr2      & 813139  & 0.307  & 0.194  & 0.190  & 0.310  & 0.975    & 1.046    & 1.054    & 0.956  & 0.993      & 0.993  \\
 chr3      & 316613  & 0.312  & 0.197  & 0.188  & 0.303  & 0.942    & 1.248    & 1.025    & 0.960  & 0.982      & 0.942  \\
 chr4      & 1531929 & 0.311  & 0.189  & 0.190  & 0.310  & 0.994    & 1.129    & 1.070    & 0.951  & 0.998      & 0.975  \\
 chr5      & 576869  & 0.306  & 0.190  & 0.195  & 0.309  & 0.941    & 1.024    & 1.030    & 0.969  & 0.992      & 0.991  \\
 chr6      & 270148  & 0.307  & 0.193  & 0.194  & 0.306  & 0.948    & 1.128    & 0.980    & 1.007  & 0.998      & 0.949  \\
 chr7      & 1090937 & 0.310  & 0.190  & 0.190  & 0.309  & 0.955    & 1.013    & 1.011    & 0.962  & 0.999      & 0.998  \\
 chr8      & 562639  & 0.309  & 0.194  & 0.191  & 0.306  & 0.990    & 1.024    & 1.162    & 0.959  & 0.994      & 0.959  \\
 chr9      & 439885  & 0.305  & 0.194  & 0.195  & 0.306  & 0.973    & 1.006    & 1.161    & 0.984  & 0.998      & 0.960  \\
 chr10     & 745444  & 0.310  & 0.191  & 0.193  & 0.306  & 0.971    & 1.024    & 1.128    & 0.974  & 0.994      & 0.974  \\
 chr11     & 666445  & 0.309  & 0.192  & 0.189  & 0.310  & 0.972    & 1.030    & 1.113    & 0.968  & 0.996      & 0.979  \\
 chr12     & 1078173 & 0.307  & 0.193  & 0.192  & 0.309  & 0.966    & 1.090    & 1.099    & 0.971  & 0.997      & 0.997  \\
 chr13     & 924430  & 0.310  & 0.191  & 0.191  & 0.308  & 0.963    & 1.050    & 1.044    & 0.948  & 0.998      & 0.995  \\
 chr14     & 784328  & 0.308  & 0.193  & 0.193  & 0.306  & 0.984    & 1.072    & 1.200    & 0.975  & 0.998      & 0.968  \\
 chr15     & 1091284 & 0.311  & 0.192  & 0.190  & 0.307  & 0.972    & 1.074    & 1.054    & 0.958  & 0.994      & 0.992  \\
 chr16     & 948061  & 0.310  & 0.190  & 0.190  & 0.309  & 0.950    & 1.025    & 0.986    & 0.987  & 0.999      & 0.981  \\
\end{tabular}
\end{ruledtabular}
\end{table*}

\begin{table*}
\caption{\label{tab:hsap} Information of the \textit{Homo sapiens}
genome. Contigs used for analysis is list in the column of
``contig" and their length is also listed. }
\begin{ruledtabular}
\begin{tabular}{lcccccccccccc}
Chromosome &Contig &Length     & A$\%$  & C$\%$  & G$\%$  & T$\%$  &$\beta_A$ &$\beta_C$ &$\beta_G$ &$\beta_T$ &$S_{base}^{1}$ &$S_{beta}^{1}$\\
\colrule
 chr1      & 30    &  36790572 & 0.293  & 0.207  & 0.207  & 0.293  & 0.972    & 1.148    & 1.080    & 0.971   & 1.000    & 0.983      \\
 chr2      &  5    &  84213153 & 0.306  & 0.193  & 0.194  & 0.307  & 0.962    & 1.073    & 1.080    & 0.965   & 0.998    & 0.998      \\
 chr3      &  2    & 100530261 & 0.305  & 0.195  & 0.195  & 0.305  & 0.962    & 1.198    & 1.097    & 0.959   & 1.000    & 0.975      \\
 chr4      &  8    &  62915881 & 0.314  & 0.185  & 0.186  & 0.315  & 0.945    & 1.098    & 1.070    & 0.942   & 0.998    & 0.992      \\
 chr5      &  8    &  41199371 & 0.307  & 0.193  & 0.192  & 0.307  & 0.947    & 1.023    & 1.037    & 0.946   & 0.999    & 0.996      \\
 chr6      &  6    &  61695806 & 0.308  & 0.192  & 0.191  & 0.309  & 0.966    & 1.080    & 1.083    & 0.964   & 0.998    & 0.999      \\
 chr7      &  5    &  64412912 & 0.304  & 0.196  & 0.196  & 0.304  & 0.952    & 1.094    & 1.104    & 0.950   & 1.000    & 0.997      \\
 chr8      &  2    &  48689376 & 0.306  & 0.194  & 0.194  & 0.305  & 0.969    & 1.082    & 1.074    & 0.972   & 0.999    & 0.997      \\
 chr9      &  1    &  39435726 & 0.306  & 0.195  & 0.194  & 0.305  & 0.952    & 1.102    & 1.045    & 0.956   & 0.998    & 0.985      \\
 chr10     &  9    &  43027086 & 0.292  & 0.207  & 0.207  & 0.294  & 0.982    & 1.077    & 1.092    & 0.971   & 0.998    & 0.994      \\
 chr11     &  2    &  48854501 & 0.296  & 0.204  & 0.204  & 0.297  & 0.954    & 1.053    & 1.052    & 0.957   & 0.999    & 0.999      \\
 chr12     &  8    &  38627316 & 0.301  & 0.200  & 0.200  & 0.300  & 0.955    & 1.014    & 1.024    & 0.948   & 0.999    & 0.996      \\
 chr13     &  3    &  67740325 & 0.310  & 0.191  & 0.190  & 0.309  & 0.955    & 1.083    & 1.119    & 0.951   & 0.998    & 0.990      \\
 chr14     &  1    &  87191216 & 0.294  & 0.204  & 0.205  & 0.297  & 0.955    & 1.154    & 1.113    & 0.956   & 0.996    & 0.990      \\
 chr15     &  2    &  22003156 & 0.291  & 0.211  & 0.210  & 0.288  & 0.958    & 1.114    & 1.085    & 0.983   & 0.996    & 0.987      \\
 chr16     &  1    &  53619965 & 0.289  & 0.211  & 0.211  & 0.289  & 0.983    & 1.091    & 1.068    & 0.981   & 1.000    & 0.994      \\
 chr17     &  5    &  24793602 & 0.282  & 0.218  & 0.218  & 0.283  & 0.951    & 1.075    & 1.175    & 0.962   & 0.999    & 0.973      \\
 chr18     &  3    &  33548238 & 0.303  & 0.197  & 0.197  & 0.302  & 0.981    & 1.075    & 1.041    & 0.984   & 0.999    & 0.991      \\
 chr19     &  1    &  31383029 & 0.262  & 0.237  & 0.238  & 0.263  & 0.971    & 1.067    & 1.060    & 0.959   & 0.998    & 0.995      \\
 chr20     &  3    &  26259569 & 0.289  & 0.209  & 0.209  & 0.293  & 0.992    & 1.093    & 1.061    & 0.980   & 0.996    & 0.989      \\
 chr21     &  1    &  28602116 & 0.306  & 0.196  & 0.195  & 0.303  & 0.971    & 1.128    & 1.289    & 0.964   & 0.996    & 0.961      \\
 chr22     &  3    &  23178213 & 0.263  & 0.237  & 0.237  & 0.263  & 0.968    & 1.050    & 1.021    & 0.993   & 1.000    & 0.987      \\
 chrX      &  9    &  32736268 & 0.304  & 0.195  & 0.195  & 0.307  & 0.973    & 1.205    & 1.167    & 0.969   & 0.997    & 0.990      \\
 chrY      &  1    &   9938763 & 0.304  & 0.194  & 0.197  & 0.305  & 0.925    & 1.199    & 1.152    & 0.913   & 0.996    & 0.986      \\
\end{tabular}
\end{ruledtabular}
\end{table*}

\begin{table*}
\caption{\label{tab:model} Information of random and simulated by
the minimal model genome.}
\begin{ruledtabular}
\begin{tabular}{lccccccccccc}
Chromosome   &Length    & A$\%$  & C$\%$  & G$\%$  & T$\%$ &$\beta_A$ &$\beta_C$ &$\beta_G$ &$\beta_T$ &$S_{base}^{1}$ &$S_{beta}^{1}$\\
\colrule
 Random      &10000000  & 0.250  & 0.250  & 0.250  & 0.250 & 0.994    & 0.992    & 0.993    & 0.992    & 1.000        & 0.999  \\
 Simulation  &1028001   & 0.270  & 0.242  & 0.249  & 0.239 & 1.004    & 0.986    & 0.979    & 0.988    & 0.962        & 0.994  \\
\end{tabular}
\end{ruledtabular}
\end{table*}

\end{document}